\documentclass[12pt,a4paper]{article}            
 \usepackage[skins,theorems]{tcolorbox}
\tcbset{highlight math style={enhanced,
  colframe=red,colback=white,arc=0pt,boxrule=1pt}}
  \usepackage[bookmarksopen, bookmarksnumbered, bookmarksopenlevel=2]{hyperref}
  \usepackage{tikz}
  \usepackage{tikz-3dplot}
 \usetikzlibrary{calc}
 \usetikzlibrary{decorations} %
 \usepackage[UKenglish]{babel}
 \usepackage[toc,page]{appendix}
 \usepackage{amsmath}
 \usepackage{amssymb}
 \usepackage{graphicx}
 \usepackage{hhline}
 \usepackage[bf]{caption}
\usepackage{cite}
\usepackage[vcentermath]{youngtab}
\usepackage{geometry}
\usepackage{slashed}
\usepackage{color}
\usepackage{stackrel}
\usepackage{tikz-cd} 
\usepackage{empheq}
\usepackage{arydshln}

\usepackage{multirow}

\newcommand{\bqa}{\begin{eqnarray}}
\newcommand{\eqa}{\end{eqnarray}}

\hypersetup{
    pdftitle={},
    pdfauthor={},
    pdfsubject={}
}
\numberwithin{equation}{section}
\numberwithin{table}{section}

\makeatletter

\newtheorem{theorem}{Claim}

\newcommand{\be}{\begin{equation}}
\newcommand{\ee}{\end{equation}}
\newcommand{\beq}{\begin{equation}}
\newcommand{\eeq}{\end{equation}}
\newcommand{\ba}{\begin{aligned}}
\newcommand{\ea}{\end{aligned}}

\newcommand{\bea}{\begin{eqnarray}}
\newcommand{\eea}{\end{eqnarray}}

\newcommand{\cV}{\mathcal{V}}

\def\unit{{1\kern-.65ex {\rm l}}}
\def\1{{1\kern-.65ex {\rm l}}}

\newcount\hour \newcount\minute
\hour=\time \divide \hour by 60
\minute=\time
\def\now{%
\ifnum \hour<13
  \ifnum \hour=0 \advance \hour by 12 \number\hour:\else \number\hour:\fi%
     \ifnum \minute<10 0\fi%
     \number\minute%
\ A.M.%
\else \advance \hour by -12 \number\hour:%
  \ifnum \minute<10 0\fi%
  \number\minute%
  \ P.M.%
\fi%
}

\def\fnote#1#2{\begingroup\def\thefootnote{#1}\footnote{#2}
     \addtocounter{footnote}{-1}\endgroup}

\makeatother

\begin{document}

\begin{flushright}
 {CERN-TH-2019-044}\\
\end{flushright}

\vskip 40 pt
\begin{center}
{\large \bf Emergent Strings, Duality and \vspace{2mm}\\Weak Coupling Limits for Two-Form Fields
} 

\vskip 11 mm

Seung-Joo Lee${}^{1}$, Wolfgang Lerche${}^{1}$, Timo Weigand${}^{1,2}$

\vskip 11 mm

\small ${}^{1}${\it CERN, Theory Department, \\ 1 Esplande des Particules, Geneva 23, CH-1211, Switzerland} \\[3 mm]
% \small ${}^{2}${\it Institut f\"ur Theoretische Physik, Ruprecht-Karls-Universit\"at, \\
% Philosophenweg 19, 69120 Heidelberg, Germany}
\small ${}^{2}${\it PRISMA Cluster of Excellence and Mainz Institute for Theoretical Physics, \\
Johannes Gutenberg-Universit\"at, 55099 Mainz, Germany}

\fnote{}{seung.joo.lee, wolfgang.lerche,  
timo.weigand @cern.ch}

\end{center}

\vskip 7mm

\begin{abstract}

We systematically analyse weak coupling limits for 2-form tensor fields in the presence of gravity. Such limits are significant for testing various versions of the Weak Gravity and Swampland Distance Conjectures, and more broadly,
the phenomenon of emergence. The weak coupling limits for 2-forms 
correspond to certain infinite-distance limits in the moduli space of string compactifications, 
where asymptotically tensionless, solitonic strings arise. These strings are identified as
weakly coupled fundamental strings in a dual frame, which makes the idea of emergence manifest.
Concretely we first consider weakly coupled tensor fields in
six-dimensional compactifications of F-theory, where the arising tensionless strings play the role of
dual weakly coupled heterotic strings. As the main part of this work, we consider
certain infinite distance limits of Type IIB strings on K3 surfaces, for which we show that the asymptotically
tensionless strings describe dual fundamental Type IIB strings, again on K3 surfaces.  
 By contrast the analogous weak coupling limits of M-theory compactifications are found to correspond to an F-theory limit where an extra dimension emerges rather than tensionless strings.
We comment on extensions of our findings to four-dimensional compactifications.
\end{abstract}

\vfill

\thispagestyle{empty}
\setcounter{page}{0}
\newpage

\tableofcontents

%%%%%%%%%%%%%%%%%%%%%%%%%%%%%%%%%%%%%%%%%%%%%%%%%
\section{Introduction and Summary}
%%%%%%%%%%%%%%%%%%%%%%%%%%%%%%%%%%%%%%%%%%%%%%%%%

Within the web of recent Quantum Gravity Conjectures, reviewed in \cite{Brennan:2017rbf,Palti:2019pca},
 the Swampland Distance Conjecture \cite{Ooguri:2006in} plays a central role.
It postulates that as one approaches a point at infinite distance in the moduli space  of any theory of quantum gravity, a tower of infinitely many states becomes asymptotically massless.
The resulting breakdown of the original effective theory   explains microscopically why the critical point is located at infinite distance in moduli space.
In the context of a gauge theory,
 the infinite tower of states becoming light in the weak coupling limit includes states that satisfy the Weak Gravity Conjecture \cite{ArkaniHamed:2006dz}, thus linking the two conjectures quite directly \cite{Heidenreich:2016aqi,Montero:2016tif,Klaewer:2016kiy} (see also \cite{Heidenreich:2015nta,Andriolo:2018lvp}).  
This has been confirmed by detailed investigations of the towers of massless states appearing at infinite distance in string theory \cite{Grimm:2018ohb,Lee:2018urn,Lee:2018spm,Grimm:2018cpv,Corvilain:2018lgw,Lee:2019tst,Joshi:2019nzi,Marchesano:2019ifh,Font:2019cxq}. The arguments do in general not rely on BPS properties of the particles \cite{Lee:2018urn,Lee:2018spm,Lee:2019tst} and hence even apply 
to theories with $N=1$ supersymmetry in four dimensions, as analyzed in \cite{Gonzalo:2018guu,Lee:2019tst}.
Refined versions of the Swampland Distance Conjecture \cite{Klaewer:2016kiy, Baume:2016psm}, examined further in string theory in \cite{Blumenhagen:2018nts}, have potentially important consequences for cosmological models relying on super-Planckian excursions of scalar fields.
%This connection has been explored in detail in the context of string theory \cite{Heidenreich:2017sim,Heidenreich:2018kpg,Grimm:2018ohb, Lee:2018urn,Lee:2018spm,Lee:2019tst}.
Emergent towers of states have been proposed in \cite{Ooguri:2018wrx} as a rationale behind the recent de Sitter conjectures \cite{Obied:2018sgi} at weak coupling.
An extensive account of the role of such towers of states, along with a survey of the recent literature, can be found in \cite{Palti:2019pca}.

A celebrated result in string theory \cite{Strominger:1995cz} is the logarithmic divergence of gauge couplings in four-dimensional string compactifications, 
which arises from a {\it finite} number of states becoming massless at special {\it finite} distance points  in moduli space.
By contrast, in a situation at infinite distance, an infinite number of states is supposedly becoming asymptotically massless. As a result the logarithmic
divergence can turn into a polynomial one.
%, which is the hallmark of a tensionless string.\footnote {At finite distance, such polynomial behaviour in four dimensions signifies a superconformal fixed point, but this is not what we consider here since we are interested in theories with gravity.}
This phenomenon is tied to the idea of emergence \cite{Harlow:2015lma,Heidenreich:2017sim,Heidenreich:2018kpg,Grimm:2018ohb,Palti:2019pca}: The fields whose inverse couplings diverge at the infinite distance point are understood as a collective phenomenon associated with the appearance of a tower of massless states. 
% Rather than integrating out these fields as we pass from the the ultra-violet to the weakly coupled infra-red, 
% we can think of
Specifically, for a gauge field that becomes weakly coupled at infinite distance, one may in principle successively integrate in the infinite tower of states as one passes to higher and higher energy scales.   The effect is to make the gauge field eventually strongly coupled and  disappear as a fundamental field at high energies \cite{Palti:2019pca}. Such behaviour by itself is not limited to four dimensions.

What has been studied in great detail in the literature up to this point is the infinite distance regime in moduli space near the weak coupling limit of some gauge field, which is a 1-form potential. 
The resulting nearly massless tower of states a priori refers to particles that are charged under this gauge field and contribute to the renormalization of the gauge coupling.  This manifestation of emergence is tied to a gauge subsector and the respective charged particles.

In this note we widen the scope of emergence and systematically analyze the weak coupling regime for 2-form gauge potentials. 
The most natural setting for such theories is in six dimensions, which is the focus of this work.\footnote{In four dimensions, 2-forms are dual to axionic fields, and Quantum Gravity Conjectures such as generalizations of the Weak Gravity Conjecture and others may have important implications e.g. for the structure of instanton corrections \cite{Rudelius:2014wla,delaFuente:2014aca,Rudelius:2015xta,Montero:2015ofa,Brown:2015iha,Bachlechner:2015qja,Hebecker:2015rya,Brown:2015lia,Heidenreich:2015wga,Kooner:2015rza,Ibanez:2015fcv,Hebecker:2015zss,Baume:2016psm,Heidenreich:2016jrl,Blumenhagen:2017cxt,Valenzuela:2017bvg,Ibanez:2017vfl,Aldazabal:2018nsj,Blumenhagen:2018nts,Blumenhagen:2018hsh,Reece:2018zvv,Hebecker:2017uix,Montero:2017yja}. The recent \cite{Marchesano:2019ifh} studies such instanton corrections in the context of emergence and the Swampland Distance Conjecture.}
Since 2-forms couple to strings, the tower of states associated with them are not just certain charged fields, but the whole string excitation spectrum.
Within the framework of string theory, six-dimensional theories with 2-forms can be constructed by compactifying Type IIB strings on complex 2-folds, denoted by $B_2$.
If $B_2$ is a K3 surface, the theory preserves 16 supercharges and corresponds to an effective $N=(2,0)$  supergravity theory.  On suitably curved K\"ahler surfaces, extra 7-branes are needed for consistency and supersymmetry  is broken to $N=(1,0)$.
Models of the latter type fall into the framework of F-theory \cite{Vafa:1996xn,Morrison:1996na,Morrison:1996pp}. By duality, the latter type of theories may also
 be described in terms of heterotic strings, but our starting point will be the Type IIB/F-Theory perspective.

One of our main results is that for such six-dimensional Type IIB or F-Theory theories, any weak coupling limit of some 2-form gauge potential leads to a tensionless solitonic string which couples precisely to that weakly coupled 2-form. As we will see,
this string corresponds to a weakly coupled, {\it critical}  fundamental string. 
The weak coupling limit of the 2-form hence forces, in a sense, the change to a new duality frame, in terms of which most excitations of the original perturbative string decouple from the low energy spectrum, while the excitations of the new fundamental string become light.
The degenerations required to take the weak coupling limit occur on the boundary of the K\"ahler moduli space.\footnote{For K3 this is true as long as we focus on {\it attractive} K3s as will be explained in section  
  \ref{sec_emergentIIB}.} As we will show, the weak coupling limit of the 2-form potential inevitably requires the shrinking of a curve of vanishing self-intersection. The asymptotically tensionless string then arises as a soliton obtained by wrapping a D3-brane along this shrinking cycle.

%In compactifications with 8 supercharges, the tensionless critical string turns out to be the heterotic string, which has already played a central role in the analysis of the Weak Gravity Conjecture for 1-form fields in our previous work \cite{Lee:2018urn,Lee:2018spm}. On the other hand, in the current work we will find that
%for Type IIB compactifications with 16 supercharges, that the tensionless solitonic string  is dual to the fundamental Type IIB string again, probing  some
%elliptically fibered K3 surface. Thus, in some sense, the theory reproduces itself at large distance, upon a change of 
%duality frame.
%

%This is reminiscent of the standard S-duality of Type IIB theory, but
% here it involves a solitonic string arising from a wrapped D3-brane. We thus may understand the weakly coupled frame as a result of the following chain of dualities:  two T-dualities followed by an S-duality.

As mentioned above, the objects which become mass- or tensionless in such weak coupling limits are not just strings per se, but these come together with the entire tower of their particle excitations (as viewed from the dual, perturbative frame).
The only other potential source of nearly massless states are Kaluza-Klein excitations of these string states, 
which may arise if in the infinite distance limit some submanifold of the compactification space simultaneously becomes large.
We will show that this in fact happens generically;
the Kaluza-Klein scale of these states turns out to be linked to the new fundamental string scale, but it is relatively suppressed by the volume ${\cal V}$ of the internal space $B_2$.
To the extent that the volume is kept fixed so as to keep gravity dynamical,
both towers of states become massless in the same large distance limit.
Thus for fixed ${\cal V}$, there is no regime where a Kaluza-Klein tower is the only tower becoming light,
 and in this sense no new dimensions open up in the effective field theory. 
The appearance of a tensionless string is crucial for this behaviour. 

Our observations appear interesting also from the perspective of the purported phenomenon of emergence: 
The computational challenge behind providing evidence for this proposal is to show how the tower of light states reproduces the observed singularities in the couplings of the effective theory at infinite distance. This has been demonstrated qualitatively in various explicit string realizations \cite{Heidenreich:2017sim,Heidenreich:2018kpg,Grimm:2018ohb,Lee:2018urn,Grimm:2018cpv,Corvilain:2018lgw},  without detailed knowledge of the contributions of the nearly massless states at each mass level. On the other hand,
for the types of large distance limits that we will discuss in this note,  emergence is manifest: The perturbative effective theory at large distance is by definition the result of integrating out the full tower of fundamental string excitations and of Kaluza-Klein excitations. In the weak coupling limit, non-perturbative effects play no role. The states which become light are precisely the fundamental string excitations plus their Kaluza-Klein tower in the new duality frame - be it the dual heterotic or the dual Type IIB frame.
Integrating out these states is guaranteed, by construction, to exactly reproduce the effective supergravity couplings. 

To appreciate the role of the emergent strings, it is fruitful to compare the situation for
 Type IIB/F-theory to the corresponding situation in Type IIA/M-theory. The analogous question to study is the weak coupling limit for 1-form gauge potentials of M-theory on K3. The limit is again controlled by the K\"ahler parameters, at least as long as we focus on attractive K3s. In such cases the weak coupling limit will be found to be identical to taking the F-theory limit.  More
specifically, the tower of asymptotically massless states are now particles (with no strings attached), which arise from M2-branes that wrap a shrinking curve. As we will show, the shrinking curve corresponds to a genus-one fiber within the attractive K3. This is a consequence of the geometry of the weak coupling limit even without further assumptions on the details of an elliptic fibration structure.\footnote{The F-theory limit of shrinking fiber volume for elliptically fibered Calabi-Yau 3-folds has been studied in \cite{Corvilain:2018lgw} as a special case within a more general analysis of infinite distance singularities in K\"ahler moduli space.}
 Interpreting these states as Kaluza-Klein states of a dual F-theory on $S^1$, one encounters an extra dimension unfolding in the weak coupling limit.
Interestingly, the unfolding of an extra dimension has also been argued in \cite{Gonzalo:2019gjp} to be a consequence of a strong version of the scalar gravity conjecture \cite{Palti:2017elp}.
By contrast, for the Type IIB version of the K\"ahler moduli degeneration we consider here, where certain 2-forms become weakly coupled in the presence of gravity and
tensionless strings emerge,  the large distance limit is physically very different and, in particular, no extra dimension opens up, in the sense described above.

In summary, the main results of our analysis of weak coupling limits of 2-forms are as follows:
\begin{itemize}
\item 
In Section \ref{sec_emergenthet} we will show that in $N=(1,0)$ supersymmetric F-theory compactifications to six dimensions, tensionless heterotic strings emerge in the
weak coupling limit of 2-forms. This is closely related to our previous work  \cite{Lee:2018urn,Lee:2018spm} which was aimed at weakly coupled gauge fields.
\item 
The main body of our work is in Section \ref{sec_emergentIIB}, 
where we consider six-dimensional compactifications of Type IIB strings with $N=(2,0)$ supersymmetry.
Out of the five possible large distance limits (two of which are trivial), we will focus on one particular of such limits, 
and find that the tensionless string that arises is again dual to the Type IIB string,
probing some elliptically fibered K3 surface. Thus, in some sense, the theory reproduces itself at large distance, upon a change of 
duality frame. This yields a six-dimensional analog of the famous picture of the moduli space of 10-dimensional M-theory, where
all large distance limits correspond to emerging weakly coupled strings (except, of course, for 11-dimensional supergravity).
This is consistent with the intuition that a gravitational theory with a weakly coupled 2-form must be a theory of critical, fundamental strings.
\item 
In Section \ref{sec_emergentF} we contrast the previous analysis with M-Theory compactifications to seven dimensions.
Here we see that in the respective large distance limits, 
towers of asymptotically massless particles arise, which are associated with an emerging extra dimension rather than with a tensionless string in the same number of dimensions.

\item
A similar picture of an emerging {\it fundamental} string governs at least some of the weak coupling limits also for F-theory/Type IIB compactifications to four dimensions, as we briefly sketch
in section \ref{sec_concl}.

\end{itemize}

{\bf Note added:} While this work was readied for publication, we received \cite{Marchesano:2019ifh} and \cite{Font:2019cxq}, which partially overlap with our results in that they study aspects of large distance limits in relation to tensionless strings and instantons in four-dimensional string compactifications.

%%%%%%%%%%%%%%%%%%%%%%%%%%%%%%%%%%%%%%%%%%%%%%%%%
\section{Emergent heterotic string from 6d  F-theory} \label{sec_emergenthet}
%%%%%%%%%%%%%%%%%%%%%%%%%%%%%%%%%%%%%%%%%%%%%%%%%

We begin by analyzing weak coupling limits for 2-form potentials in general six-dimensional F-theory compactifications with $N=(1,0)$ supersymmetry. 
The main result of this section can be summarized as follows:
\begin{theorem} \label{claim1}
Consider a limit where some 2-form gauge potential in F-theory compactified to six dimensions becomes asymptotically weakly coupled, while  gravity is kept dynamical.
This limit is at infinite distance in K\"ahler moduli space, and there emerges an asymptotically tensionless heterotic string, which is charged under the weakly coupled 2-form potential in question.
The weak coupling limit corresponds to a change of duality frame to the one of a perturbative heterotic string. 
% Furthermore, the 2-form potential coupling to the heterotic string  is the only weakly coupled 2-form in the limit. 
\end{theorem}

The relevant limit in K\"ahler moduli space will be found to be
identical to the weak coupling limits for 7-brane gauge theories in F-theory studied already in \cite{Lee:2018urn,Lee:2018spm}.
The new aspects of the discussion in this section refer to the relation of this degeneration limit to the weak coupling regime of 2-form fields.
It furthermore paves the way for an understanding of the weak coupling limits of Type IIB strings on K3 surfaces, which will be studied in Section \ref{sec_emergentIIB}.

%%%%%%%%%%%%%%%%%%%%%%%%%%%%%%%%%%%%%%%%%%%%%%%%%
\subsection{Geometric setup} \label{subsec_geomset}
%%%%%%%%%%%%%%%%%%%%%%%%%%%%%%%%%%%%%%%%%%%%%%%%%

Consider compactifications of F-theory 
to six dimensions \cite{Vafa:1996xn,Morrison:1996na,Morrison:1996pp}. The compactification space is given by a Calabi-Yau threefold, which
is an elliptic fibration over some K\"ahler surface $B_2$ with non-trivial anti-canonical bundle, $\bar K_{B_2}$.\footnote{
For an introduction to some of the techniques used to describe such backgrounds, we refer
 e.g., to \cite{Taylor:2011wt,Weigand:2018rez} and references therein.}
The low-energy effective theory is described by a six-dimensional, chiral $N=(1,0)$ supergravity theory.
The various 2-form gauge potentials result from the dimensional reduction of the Type IIB Ramond-Ramond 4-form, $C_4$, with respect to some basis of harmonic $(1,1)$ forms on $B_2$:
\be
C_4 = B^\alpha \wedge \omega_\alpha \,, \qquad \quad \omega_\alpha \in H^{1,1}(B_2) \,.
\ee
The dual basis $\{\omega^{\alpha}\}$ of curve classes on $B_2$, defined via
$\omega^\alpha \cdot \omega_{\beta} = \delta^{\alpha}_\beta $,
is related to $\{\omega_\alpha\}$ as
\be
\omega^\alpha = \Omega^{\alpha \beta} \omega_\beta \,.
\ee
Here $\Omega^{\alpha \beta} = \omega^\alpha \cdot \omega^\beta$
is the inverse of the intersection form 
\be
\Omega_{\alpha \beta} = \omega_\alpha \cdot \omega_\beta \equiv \int_{B_2} \omega_\alpha \wedge \omega_\beta \,.
\ee
The volume of $B_2$ is computed in terms of the K\"ahler form 
\be
J =  j^\alpha \omega_\alpha
\ee
as
\be \label{Mplfixed}
{\cal V} = \frac{1}{2} J \cdot J \,,
\ee
and throughout this article we keep ${\cal V}$ at a fixed value to ensure that gravity remains dynamical. 
The kinetic terms for the collection of 2-form fields $B^\alpha$,
\be \label{kin1}
S_{\rm kin} = \int_{\mathbb R^{1,5}} \frac{M^4_{\rm Pl}}{2} \sqrt{-g} R  -  \frac{2 \pi}{4} \int_{\mathbb R^{1,5}} g_{\alpha \beta} \, d B^\alpha \wedge \ast d B^\beta + \ldots  \qquad \quad {\rm with} \quad  M^4_{\rm Pl} ={4\pi} {\cal V} \,,
\ee
are then controlled by the following coupling matrix:
\be \label{galphabeta}
g_{\alpha \beta} = \int_{B_2} \omega_\alpha \wedge \ast \omega_\beta  = \frac{j_\alpha j_\beta}{\cal V}  - \Omega_{\alpha \beta} \,.
\ee
% Here 
% \be
% \Omega_{\alpha \beta} = \omega_\alpha \cdot \omega_\beta \equiv \int_{B_2} \omega_\alpha \wedge \omega_\beta
% \ ee
 % denotes the intersection product in $H^{1,1}(B_2)$
The pseudo-action (\ref{kin1}) follows by dimensional reduction from the 10d  action in the Einstein frame with $\ell_s \equiv 1$.
Note that (\ref{galphabeta}) is an immediate consequence of the fact that on a K\"ahler surface of volume ${\cal V}$ 
\be \label{astomega}
\ast \omega_\alpha =  \frac{J \cdot \omega_\alpha}{\cal V}  \, J - \omega_\alpha \,.
\ee
As usual for six-dimensional $N=(1,0)$ supergravity, the gauge invariant field strengths, $H^\alpha$, derived from the 2-form potentials $B^\alpha$, 
are subject to the self-duality condition
\be \label{em6d}
g_{\alpha \beta} \ast H^\beta = \Omega_{\alpha \beta} H^\beta \,,
\ee
where the field strengths involve in addition  
suitable Chern-Simons terms which are respon\-sible for the correct cancellation of local anomalies. Details can be found e.g.~in \cite{Bonetti:2011mw} and in the references therein.

The 2-form gauge potentials couple to effective strings in ${\mathbb R^{1,5}}$ which arise from D3-branes which wrap holomorphic curves on $B_2$.
To discuss this, it is useful to introduce another basis of curve classes 
\be
C^I = C^I_\alpha  \, \omega^\alpha \,,
\ee
and their dual divisors $C_J = C_J^\alpha \omega_\alpha$ such that  $C^I \cdot C_J = \delta^{I}_{J}$. We will also need the intersection form 
\be
\Omega_{IJ} = C_I \cdot C_J = C_I^\alpha   C_J^\beta \Omega_{\alpha \beta}\,,
\ee
as well its inverse, $\Omega^{IJ}$, in order to lower and raise the $I$ index.
% For instance 
% \be
% C_\beta^I = C_{J}^\alpha \,  \Omega^{IJ}  \, \Omega_{\alpha \beta} \,.
% \ee

% and to denote by $C_J^\beta$ the inverse of the matrix $C_I^\alpha$,
% \be
% C_\beta^I   \, C_I^\alpha = \delta_\beta^\alpha \,.
% \ee

In terms of these, the coupling of the 2-forms to the string obtained by wrapping a D3-brane along the curve $C^I$ derives from the D3-brane Chern-Simons couplings
 as follows:
\bea \label{BIcouplings}
S_{\rm CS} &=& 2 \pi \int_{\mathbb R^{1,1} \times C^I} C_4 = 2 \pi \int_{\mathbb R^{1,1} \times B_2} (B^\alpha \wedge \omega_\alpha)  \wedge (C^I_\beta \,  \omega^\beta) \\
%= 2 \pi \int_{\mathbb R^{1,1}} B^\alpha \mega_{\alpha \beta} C_I^\beta  \\
&=:& %2 \pi \int_{\mathbb R^{1,1}} B_I = 
2 \pi \int_{\mathbb R^{1,1}}  B^I  \,. % \equiv 2 \pi \int_{\mathbb R^{1,1}}  B_I    \,.
\eea
Here, the 2-forms $B^I$ are related to the fields $B^\alpha$
as 
% with
% \be
% B_I = B^\alpha C_{I \alpha} = B^\alpha \, C_I^\beta \, \Omega_{\alpha \beta} \,.
% \ee
% This relation can equivalently be written as
\be
B^I = C^I_\alpha B^\alpha \,.
\ee
The matrix $g_{IJ}$ of kinetic terms associated with these linear combinations $B^I$ is then obtained by equating
\be \label{skingIJ}
S_{\rm kin} =   - \frac{2\pi}{4} \int_{\mathbb R^{1,5}} g_{\alpha \beta} \, d B^\alpha \wedge \ast d B^\beta % =  \frac{1}{4} \int_{\mathbb R^{1,5}}   g_{\alpha \beta}  C_I^{\alpha}  C^{\beta}_J  d B^I \wedge \ast d B^J  = 
=  - \frac{2\pi}{4} \int_{\mathbb R^{1,5}} g_{I J} d B^I \wedge \ast d B^J,
\ee
which yields
% \be
% g^{IJ} =  (C^{-1})_{\alpha}^I  (C^{-1})_{\beta}^J g^{\alpha \beta}  \,.
% \ee
% The strength of the couplings of the $B_{I}$ is encoded in the inverse matrix
\be \label{gIJ}
g_{I J } = C_I^\alpha C_J^\beta g_{\alpha \beta} =
\frac{1}{\cal V}  (J \cdot C_I)  (J \cdot C_J)  - C_I \cdot C_J \,.
%{\rm vol}(C_I) {\rm vol}(C_J)  - C_I \cdot C_J \,, \qquad \quad {\rm vol}(C_I) = J \cdot C_I \,.
\ee
As long as the divisors $C_I$ are effective, 
the matrix $g_{I J }$ depends on the volumes of 
$C_I$ via
\be
{\rm vol}(C_I) = J \cdot C_I \,.
\ee
Note however that the expression (\ref{gIJ}) holds also for more  general divisor classes, irrespective of whether they admit a holomorphic representative. 

% Conversely, the kinetic matrix of the linear combinations  of 2-forms $B_I$ coupling to the string associated with $C_I$ is given by
% \be \label{g^IJ}
% g^{IJ} = \Omega^{IK}  \Omega^{JL} g_{KL} \,.
% \ee
% where
% \be

% \ee

%%%%%%%%%%%%%%%%%%%%%%%%%%%%%%%%%%%%%%%%%%%%%%%
\subsection{Weak coupling limit for 2-form potentials} \label{weakcouplinghet}
%%%%%%%%%%%%%%%%%%%%%%%%%%%%%%%%%%%%%%%%%%%%%%%

With this preparation we can now analyse the possible weak coupling limits for the 2-form potentials, $B^I$, under the side condition that gravity is not decoupled.
On general grounds, a tensor field can only be weakly coupled if it is a linear combination of a self-dual and anti-self-dual tensor field.
Sten-dimensionalix-dimensional $N=(1,0)$ supergravity contains one self-dual tensor, which is part of the gravity multiplet, along with $n_T$ anti-self-dual tensors in tensor multiplets.
A weak coupling limit  must thus single out some linear combination of anti-self-dual tensors to combine with the gravitensor to form the desired
weakly coupled tensor field, ${\rm T}$.
In a suitable basis, its kinetic terms can be written as
\be \label{SkinTa}
S_{\rm kin} = \int_{\mathbb R^{1,5}} \frac{M^4_{\rm Pl}}{2} \sqrt{-|g|} R - \frac{2\pi}{4} \left( S^{2} \,  d{\rm T} \wedge \ast d{\rm T}   + S^{-2}\,  d \tilde {\rm T} \wedge \ast d \tilde {\rm T} \right) + \ldots    \qquad {\rm for} \quad S \to \infty \,.
\ee
Here $S$ can be interpreted as a ``dilaton'' that couples to the weakly coupled tensor field ${\rm T}$ and determines its inverse gauge coupling.
On the other hand, the dual tensor $\tilde {\rm T}$, defined by\footnote{The dots stand for the Chern-Simons corrections to the field strength, which play no role for us. They will be omitted in the sequel.}
\be  \label{SkinTb}
 d  ({\rm T} + \ldots) = S^{-2} \ast d (\tilde {\rm T} + \ldots),
\ee
 becomes strongly coupled in the limit $S \to \infty$.  
Note that in this limit, the remaining anti-self-dual tensors do not mix with ${\rm T}$ nor with $\tilde {\rm T}$.

To identify such a limit for the system of tensor fields in (\ref{skingIJ}), we analyze  the matrix of gauge kinetic terms, $g_{IJ}$,  for the 2-form gauge fields, $B^I$. In order for a weak coupling limit for (at least) one linear combination of potentials to be attained, (at least) one eigenvalue of $g_{I J }$ must tend to infinity. At the same time, we must keep
 the volume of the base, $B_2$, fixed as in (\ref{Mplfixed}), in order to keep gravity dynamical.
 This means that there must exist at least one divisor class, $C$, for which 
 \be \label{JCinft}
 J \cdot C \to \infty \,, \quad {\rm with} \qquad \frac{1}{2} J \cdot J = {\cal V} \quad {\rm fixed}\,.
 \ee 
This type of geometric degeneration limit coincides precisely with the type of infinite distance limits that were 
analyzed in full generality in refs.~\cite{Lee:2018urn,Lee:2018spm}.
The motivation  of \cite{Lee:2018urn,Lee:2018spm} to study this limit was a priori independent of the goal of the present paper: 
These references consider the weak coupling limit for 1-form gauge fields localised on  7-branes that wrap certain divisors, $C$, of $B_2$.
The inverse coupling of these gauge theories is proportional to $J \cdot C$, and therefore (\ref{JCinft}) governs their weak coupling limit as well.
We thus see that the same geometric limit also governs the weak coupling regime of the 2-form potentials, whose inverse coupling matrix is given by (\ref{gIJ}).

It was shown in \cite{Lee:2018urn,Lee:2018spm} that
whenever a limit of the form (\ref{JCinft}) can be taken, the K\"ahler form must behave asymptotically as
\be \label{WCL1}
J = t J_0 + \sum_i \frac{a_i}{2 t} J_i \qquad {\rm for} \qquad t \to \infty \,,
\ee
where $J_0$ and $J_i$ are generators of the K\"ahler cone and $C \cdot J_0 \neq 0$.
Here $J_0$ is singled out as the generator of the K\"ahler cone whose coefficient scales to infinity, which  is thus responsible for the volume of some
 curve class to diverge.
As proven in \cite{Lee:2018urn},  eq.~(\ref{WCL1}) is the most general ansatz that realises the limit (\ref{JCinft}).
In order for  the base volume to remain finite  in the limit $t \to \infty$, as in (\ref{Mplfixed}),
the K\"ahler generators $J_0$ and $J_i$ must satisfy 
\be \label{J0J0zero}
J_0 \cdot J_0 = 0 \,, \qquad \quad \sum_i \frac{a_i}{2} J_i \cdot J_0 = {\cal V} + {\cal O}(1/{t^2}) \,,
\ee
where $a_i$ are parameters that stay finite in the limit $t \to \infty$. 

An important property of the asymptotic behavior (\ref{WCL1}) is that the class associated with $J_0$ necessarily 
contains a holomorphic, rational curve 
\be
C^0 := J_0 \, \qquad {\rm with} \quad C^0 \cdot C^0 = 0 \,,
\ee
whose volume vanishes in the limit as 
\be \label{volC0}
{\rm vol}(C^0) = \sum_i \frac{a_i}{2 t} J_i \cdot J_0  = \frac{\cal V}{t} + {\cal O}(1/{t^3})  \qquad {\rm for} \quad t \to \infty \,.
\ee
This implies that a D3-brane wrapping this curve $C^0$ gives rise to a solitonic string in $\mathbb R^{1,5}$ with tension
\be \label{tensionC0}
T_{C^0} = 2 \pi \,  {\rm vol}(C^0) =  2\pi \frac{{\cal V}}{t}  + {\cal O}(1/{t^3}) \,,
\ee
in the frame defined via (\ref{kin1}).

Before giving an interpretation of this solitonic string in Section \ref{subsec_emergeht},
we now show that 
it  becomes weakly coupled in the tensionless limit.
Viewing the string as the object charged under some given, definite linear combination of 2-forms, 
it becomes weakly coupled whenever the diagonal kinetic term (\ref{gIJ}) of the  relevant 2-form diverges and there is no significant kinetic mixing involving this 2-form. 
To check this explicitly, consider the following basis of curve classes 
\be \label{C^Idef}
\{C^I\} = \{C^0:= J_0, \ \  C^i := J_i\} \,.
\ee
Note that we are only using the property that $C^0$ contains a holomorphic curve class, as established in \cite{Lee:2018urn}, while the remaining
classes, $C^i$, need not have this property.
We are interested in the kinetic terms for the linear combination $B^0$ of 2-forms that couples to $C^0$ as shown  in 
 (\ref{BIcouplings}).
In particular, we need to analyze the limit $t \to \infty$ of the kinetic terms $g_{00}$ and $g_{0i}$,  where $g_{IJ}$ is defined in (\ref{gIJ}).

% As we will see by analyzing the behaviour of the kinetic matrix $g^{IJ}$, this combination of 2-forms is indeed weakly coupled in the limit $t \to \infty$. 

To this end, introduce the following dual basis of divisors $C_I$ which has the properties
\be \label{dualprops}
C_0 \cdot C^0 = 1 \,, \qquad C_i \cdot C^0 = 0 \,, \qquad   C_0 \cdot C^i = 0 \,, \qquad C_i \cdot C^j = \delta_i^j  \,.
\ee
In the limit (\ref{WCL1}),
\bea
J \cdot C_0 &=&  t J_0 \cdot C_0 + \sum_i \frac{a_i}{2 t} J_i \cdot C_0 = t \,, \\
J \cdot C_i  &=&  t J_0 \cdot C_i + \sum_j \frac{a_j}{2 t} J_j \cdot C_i =  \frac{a_i}{2t} \,,
\eea
where we used (\ref{C^Idef}).
As a result, the metric (\ref{gIJ}) becomes
\bea \label{g00gijasmp}
g_{00} &=& \frac{t^2}{\cal V}  - C_0 \cdot C_0  \,, \qquad g_{0 i } = \frac{a_i}{2 \cal V} - C_0 \cdot C_i \,  \label{g001}\\
g_{ij} &=& \frac{a_i a_j}{4 {\cal V} t^2} - C_i \cdot C_j   \label{g002} \,.
\eea
In the limit $t \to \infty$, the mixing of the 2-form $B^0$ with the remaining 2-forms 
becomes negligible in the sense that $g_{0i}/g_{00} \to 0$. Together with the fact that $g_{00} \sim t^2 \to \infty$, this guarantees that the string associated with $C^0$ becomes weakly coupled.
Importantly, the remaining kinetic couplings, $g_{ij}$,  stay finite in the limit $t \to \infty$, and do not exceed values of order one.
This means that in the specific basis we have chosen, $\{\omega_\alpha \} = \{J_0, C_i\}$, 
$B^0$ is the {\it only} 2-form gauge field that becomes weakly coupled as $t \to \infty$, while all other linear combinations of fields $B^i$ become strongly coupled.

To make contact with (\ref{SkinTa}) and  (\ref{SkinTb}),
we fist recall that the split of 2-form potentials into self-dual and anti-self-dual tensors can be found by diagonalizing the duality matrix,\footnote{A detailed discussion of the split of 2-form potentials into self-dual and anti-self-dual tensors can be found in Section~$2$ of~\cite{Lee:2018ihr}.} 
\beq
D^{I}_{~\,J} :=  (g^{-1})^{IK} \Omega_{K J} = \frac{1}{\cV} (J \cdot C^I) (J \cdot C_J) - \delta^I_{~\,J}\,,
\eeq
which obeys 
\beq
*H^I = D^I_{~\,J} H^J \,. 
\eeq
Then, in an appropriate basis of curves, $\{C'^I\}$, the duality matrix can be taken in a diagonal form,
\beq
D'^I_{~\,J} = {\rm diag}(+1, -1, -1, \cdots, -1)\,.
\eeq
%for which the associated $2$-form fields $\{B'^I\}$ are $B^{(+)}$, $B^{(-)}$, and $B^{(-)}_{i=2, \cdots, n_T}$. 
%Here, the prime indicates that the physical quantities in question are the ones uniquely determined by the choice $C'^I$ of the basis curves. 
Upon a further change of basis % that mixes the $B^{(+)}$ and $B^{(-)}$ in an appropriate manner, 
one can find another basis of curves, $\{C''^I\}$, with respect to which the duality matrix takes the form
\begin{eqnarray}\label{B-basis}
D''^I_{~~J}= \def\arraystretch{1.15}\left(\begin{array}{cc|ccc} 
0 & S^{-2} & 0 & \cdots &0 \\ 
S^2 & 0 & 0 & \cdots & 0 \\ \hline
0 & 0 & -1 & \cdots & 0 \\
\vdots & \vdots & \vdots & \ddots&\vdots\\
0 & 0 & 0 & \cdots& -1\\ 
\end{array}\right)  \ .
\end{eqnarray}
Correspondingly, the set of $2$-forms, $\{B''^I\}$, comprises ${\rm T}$, $\tilde {\rm T}$, plus  the remaining anti-self-dual 2-forms, $B^{(-)}_{i=2, \cdots, n_T}$. 
% Note that the off-diagonal entries $\gamma$ and $1/\gamma$ may not equal to $1$ and set how the magnetic dual of $\rm B$ is defined in terms of its Hodge dual,{\footnote{This amounts to the residual freedom in normalization of the $\tilde {\rm B}$-field kinetic term in the $6$-dimensional effective action. There is of course another source of arbitrariness, already in the $10$-dimensional string frame action, due to the division of the kinetic terms for Kalb-Ramond $2$-form and its magnetic dual $6$-form. We fix the latter by taking the democratic action.}}
% \beq
% *d {\rm B} = \gamma\, d \tilde {\rm B} \,.
% \eeq
The form of (\ref{g00gijasmp}) guarantees that  in this latter basis, 
\be
C''^0 = C^0 + {\cal O}(1/t)  := J_0 + {\cal O}(1/t) \,.
\ee 
We can hence identify the weakly coupled tensor and its associated coupling parameter as
\be
{\rm T} = B^0 + {\cal O}(1/t) \,, \qquad S =  \frac{t}{\sqrt{\cal V}} \,.  
\ee
Most importantly, 
the string from the D3-brane wrapping $C^0$ is identified with a weakly coupled string, ${\cal S}$, whose tension satisfies the relation
\be
\frac{T_{\cal S}}{M^2_{\rm Pl}} = \sqrt{\pi} S^{-1} \,.
\ee
The existence of this tensionless string in the weak coupling limit is in agreement with expectation from the Swampland Distance Conjecture.
Even though we have not explicitly constructed it in full generality, it is clear that the D3-brane on the curve $C''^1$ gives rise to a dual string, $\tilde S$, with tension
\be
\frac{T_{\tilde{\cal S}}}{M^2_{\rm Pl}}  = \sqrt{\pi} S \,.
\ee
This can be confirmed by investigating explicit examples.

%%%%%%%%%%%%%%%%%%%%%%%%%%%%%%%%%%%%%%%%%%
\subsection{Emergent heterotic string at weak coupling} \label{subsec_emergeht}
%%%%%%%%%%%%%%%%%%%%%%%%%%%%%%%%%%%%%%%%%%

As discussed in our previous work \cite{Lee:2018urn}, the correct interpretation of the tensionless string is that it takes the role of the weakly coupled heterotic string in a new duality frame, that is:

\begin{equation}
\boxed{
\label{hatgsh}
\text{F-theory on $B_2$ }  \qquad  \stackrel{\text{limit} \, (\ref{WCL1}), (\ref{J0J0zero})}{\Longrightarrow}   \qquad   \text{Heterotic on K3 with}   \,    S^2_{\rm het} := {\cal V}_{\rm het} e^{-\Phi_{\rm het}} = \frac{t^2}{\cal V} \,.}
% \end{center}
\end{equation}

Indeed, an analysis of the $N=(0,4)$ supersymmetric worldsheet theory of the solitonic string  from the D3-brane wrapped on $C^0$ identifies it with
the critical heterotic string compactified to six dimensions \cite{Lee:2018urn} on a certain K3 surface.
 This realizes the well-known phenomenon that the critical heterotic string can be viewed as a soliton \cite{Harvey:1995rn} and serves as the microscopic rationale behind standard F-theory-heterotic duality \cite{Vafa:1996xn,Morrison:1996na,Morrison:1996pp}.
 The novel insight of our analysis is that a weak coupling limit for any 2-form in F-theory inevitably realises a version of this duality:
In the limit (\ref{WCL1}) the theory switches duality frame from the original Type IIB/F-theory frame to a dual heterotic frame,
 where the role of the fundamental string is played by the solitonic string associated with $C^0$.

This implies that in the duality frame defined by the weakly coupled string, the specific linear combination ${\rm T} = B^0$ of 2-forms to which this string couples plays the role of the ``universal'' 2-form Kalb-Ramond field ${\rm B}$ that couples to the fundamental string. 
% Recall that for the heterotic string, the kinetic term of the universal B-field is controlled by the heterotic dilaton $S_{\rm het}$. 
% More precisely, let us start from the ten-dimensional action in the string frame,
% \be
% S = \frac{2\pi}{\ell_{\rm het}^8} \int_{\mathbb R^{1,9}} \left(  e^{-2\Phi_{\rm het}} \sqrt{-g} R - \frac{1}{2} e^{-2\Phi_{\rm het}}   d {\rm B} \wedge \ast d {\rm B} \right) \,,
% \ee
% and perform the standard Weyl rescaling 
% \be \label{Weylrescaling}
% g_{MN} \rightarrow e^{\frac{4}{d-2} \, \Phi_{\rm het}} g_{MN}  \,, \qquad \text{with} \quad  d=10 
% \ee
% n order to switch to the ten-dimensional Einstein frame. 
To set the notation, recall that dimensional reduction to six dimensions on K3 of the heterotic 10d Einstein frame action 
(with $\ell_{\rm het} \equiv 1$) yields
\be \label{6dhetEinst}
S^{(E)}_{\rm het, 6d}=    \int_{\mathbb R^{1,5}}\left( \frac{M^4_{\rm Pl}}{2} \sqrt{-g} R - \frac{2\pi}{4}  (S^2_{\rm het}    d {\rm B} \wedge \ast d {\rm B} + S^{-2}_{\rm het}    d \tilde {\rm B} \wedge \ast d \tilde {\rm B}) \right)\,,    \,
\ee 
where 
\be
M^4_{\rm Pl} = 4 \pi {\cal V}_{\rm het} \,,  \qquad S^2_{\rm het } = {\cal V}_{\rm het} \, e^{- \Phi_{\rm het}}   \,.
\ee
We have written the kinetic terms in a democratic fashion by introducing the ``magnetically'' dual field, $\ast d \tilde {\rm B} = S^{2}_{\rm het} d {\rm B}$.
It couples to a dual string which arises as an NS5-brane wrapping the K3 on the heterotic side.
Since we have kept the physical Planck mass fixed in taking the weak coupling limit on the F-theory side, 
the volume of the dual heterotic K3 surface emerging as the weakly coupled description must equal the volume of $B_2$ on the F-theory side,
  \be
  {\cal V_{\rm het}}={\cal V} \,.
  \ee
The identification (\ref{hatgsh}) then follows by identifying the 
tension of the fundamental heterotic string in the above frame,
\be
\frac{T_{F_{\rm het}}}{M^2_{\rm Pl}} = \sqrt{\pi}  \frac{e^{\Phi_{\rm het}/2}}{\sqrt{\cal V}}  =  \sqrt{\pi}  S^{-1}_{\rm het}  \,, 
\ee
with the tension of the asymptotically weakly coupled string, ${\cal S}$, obtained on the F-theory side:
\be
S = S_{\rm het}    \qquad  \rightarrow  \qquad e^{\Phi_{\rm het}} = \frac{{\cal V}^2}{t^2}  \,.      
\ee

To summarize, whenever we take a weak coupling limit for some (linear combination of)  2-form potentials in F-theory, which we denote by $B^0 \equiv {\rm T}$, while keeping gravity dynamical, a tensionless, weakly coupled string ${\cal S}$ arises, which is dual to a heterotic string propagating in six dimensions. The 2-form potential  under which this string is charged is precisely the 2-form $B^0$ which becomes weakly coupled. 
The theory is then best described by switching the duality frame to the frame where the asymptotically weakly coupled string acts as the fundamental string, and  where $B^0$ takes the role of the heterotic Kalb-Ramond 2-form potential. 

Since in the weak coupling limit the heterotic string becomes tensionless, all its excitation modes become massless. Integrating out these modes reproduces the perturbative supergravity effective action. In this sense, the running of the coupling constants in the effective action is emergent: It can either be understood from the perspective of the original supergravity theory, or by integrating out the full tower of asymptotically light modes arising from the emergent fundamental (here, heterotic) string.

Note that in addition to the modes of the nearly tensionless string, the theory potentially contains an extra tower of Kaluza-Klein modes, which are simply the Kalzua-Klein modes of the nearly tensionless string along any submanifold that simultanously happens to become large in the limit (\ref{WCL1}).
 We will elaborate further on this point at the end of section \ref{subsec_emergent}, where we will encounter a similar emergence of a fundamental string,
 together with a Kalzua-Klein tower, in the weak coupling limit of a different theory. 
 %The KK tower is likewise integrated out automatically in the effective supergravity action describing the weak coupling regime.

%%%%%%%%%%%%%%%%%%%%%%%%%%%%%%%%%%%%%%%%%%%%%%%%%%%
\section{Emergent fundamental strings from Type IIB strings on K3} \label{sec_emergentIIB}
%%%%%%%%%%%%%%%%%%%%%%%%%%%%%%%%%%%%%%%%%%%%%%%%%%%

In this section we analyze the weak coupling limit 
for 2-form potentials of Type IIB string theory compactified on an ``attractive'' K3 surface.\footnote{We will explain this notion further below.}
Our main result is summarised in
\begin{theorem}
Consider a certain limit where a 2-form gauge potential in Type IIB string theory compactified on an attractive K3 becomes asymptotically weakly coupled, while  gravity is kept dynamical.
In this limit, there emerges an asymptotically tensionless solitonic string which is charged under the weakly coupled 2-form potential.  Via duality it corresponds
 to  a critical Type IIB string propagating on an elliptic K3 surface.
\end{theorem}
%that the weak coupling limit leads to a tensionless solitonic string which is to be identified with a fundamental Type IIB string probing the same K3. 
Thus, in a sense the theory reproduces itself 
in the duality frame relevant at infinite distance in moduli space, which is of considerable significance for the proposal of emergence  \cite{Harlow:2015lma,Heidenreich:2017sim,Heidenreich:2018kpg,Grimm:2018ohb,Palti:2019pca}.

%%%%%%%%%%%%%%%%%%%%%%%%%%%%%%%%%%%%%%%%%%%%%%%%%%%
\subsection{Large distance limits in the  moduli space ${\cal M}_{(5,21)}$} \label{sec_weakcouplSUGRA}
%%%%%%%%%%%%%%%%%%%%%%%%%%%%%%%%%%%%%%%%%%%%%%%%%%%

Type IIB string theory compactified on a K3 surface is approximated, at low energies, by an effective 6d supergravity theory with $N=(2,0)$ supersymmetry. 
Many important properties of K3 surfaces as probed by Type IIB string theory can be found in \cite{Aspinwall:1996mn} and references therein. 
The bosonic part of the gravity multiplet of 6d $N=(2,0)$ supergravity contains the metric degrees of freedom as well as five self-dual tensors $B^{(+)}_i$, $i=1, \ldots 5$, which transform as ${\bf 5}$ of the R-symmetry group $Sp(4)$.
The remaining degrees of freedom organize into tensor multiplets, each of which containing, at the bosonic level, one anti-self dual tensor along with $5$ real scalars. 
For Type IIB string theory on K3, there are altogether $21$ such tensor multiplets, into which the $105$ moduli fields organize.
This 105 dimensional, non-perturbative moduli space of Type IIB string theory on K3 can be written as the coset space
\be \label{modulispaceK31}
{\cal M}_{(5,21)} = O(\Gamma_{5,21})  \backslash O(5,21) /  O(5) \times O(21) \,,
\ee
where $O(\Gamma_{5,21})$ represents the discrete $U$-duality group of the theory.
 
The total of 26 self-dual or anti-self-dual tensor fields of the theory arise as follows:
The 10d Kalb-Ramond field $B_2$ and the Ramond-Ramond cousin $C_2$ give rise to a 6d tensor field each, which is neither self-dual nor anti-self-dual and is accompanied by its respective dual counterpart, $\tilde B_2$ and $\tilde C_2$. 
In addition, the theory contains $h^{2}(K3) = 22$ massless self-dual or anti-self-dual tensor fields, $B^A$, which arise from dimensional reduction of the Ramond-Ramond 4-form $C_4$ field via
\be \label{C4decomp}
C_4 = B^A \wedge \omega_A \, , \qquad \omega_A \in H^2(K3, \mathbb R) \,, \qquad A= 1, \ldots, 22 \,.
\ee
% If we denote the 26 self-dual/anti-dual tensors as $B^{\cal J}$, the duality transformation relating them is
% \be
% B_{\rm D}^{\cal J} = G^{\cal J}_{\cal I} B^{\cal I} \,,
% \ee 
% where in a stuiable basis 
% \be
% G^{\cal J}_{\cal I} = \begin{matrix}    \end{matrix}
% \ee

Our interest is in the general structure of the possible weak coupling limits for the tensor fields at infinite distance in moduli space.
The moduli space (\ref{modulispaceK31}) contains five non-compact directions, which we parametrize by moduli $S_i$, $i=1,\ldots, 5$.
Moving to infinite distance along any of these five directions gives rise to a generally different weak coupling limit.\footnote{Note that for taking such limits, one must specify how precisely the boundary of ${\cal M}_{(5,21)}$ is approached. We will come back to this point below.}
Each of the five non-compact scalars $S_i$ sits in a separate tensor multiplet. The associated anti-self-dual 2-forms, $B^{(-)}_i$,  $i=1,\ldots, 5$,  pair up with the $5$ self-dual tensors in the gravity multiplet, $B_i^{(+)}$, to produce a six-dimensional 2-form tensor field ${\rm T}_i$
(which is neither self-dual nor anti-self-dual),  plus its dual, $\tilde {\rm T}_i$. 
In a given weak coupling limit parametrized by $S_i$, the tensor field ${\rm T}_i$ becomes weakly coupled  (resp.~its dual, $\tilde {\rm T}_i$, in the opposite limit) .

More specifically,
let us fix our notation by taking the kinetic terms for the given weakly coupled system of tensors to be
\be
S_{\rm 6d} = \int_{\mathbb R^{1,5}} \left(\frac{1}{2} M^4_{\rm Pl} \sqrt{- |g|} R  -  \frac{2 \pi}{4}  (S_i^2 d{\rm T}_i \wedge \ast d {\rm T}_i +     S_i^{-2} d \tilde {\rm T}_i \wedge \ast d \tilde {\rm T}_i) + \ldots  \right) 
\ee
with 
\be
\ast d{\rm T}_i = S^{-2}_i  \, d \tilde {\rm T}_i \, 
\ee
and no significant kinetic mixing of the weakly coupled tensor $T_i$ with the other tensor fields in the limit ${S_i} \to \infty$.
This corresponds to the conventions
\bea
S_i \to \infty: \qquad    &{\rm T}_i \, \, \, \,  \text{becomes weakly coupled \,,}   \qquad i=1,\ldots, 5 \,, \\
S^{-1}_i \to \infty: \qquad    &\tilde {\rm T}_i \, \, \, \,  \text{becomes weakly coupled \,,}   \qquad i=1,\ldots, 5 \,.
\eea
As $S_i \to \infty$, the string ${\cal S}_i$ to which ${\rm T}_i$ couples ``electrically'' becomes tensionless, while the dual string $\tilde {\cal S}_i$, to which  $\tilde {\rm T}_i$ couples, becomes heavy.
That is, the scalar $S_i$ is related to the tension of the string ${\cal S}_i$ that couples electrically to ${\rm T}_i$ in the following way:
\bea
T_{{\cal S}_i}/M^2_{\rm Pl} &=& \sqrt{\pi} \,  S_i^{-1} \\
T_{\tilde {\cal S}_i}/M^2_{\rm Pl} &=& \sqrt{\pi} \,  S_i .
%S^{-1}_i = \frac{1}{\sq}T_{{\cal S}_i}/M^2_{\rm Pl} \,.
\eea
The division by the squared Planck mass normalizes the tension of the strings with respect to the 
six-dimensional Planck scale, and the numerical factors are chosen such that the tension of the string and its dual satisfy the canonical relation
\be
T_{{\cal S}_i} T_{\tilde{\cal S}_i}  = \pi M^4_{\rm Pl} \,.
\ee

\begin{figure}[t!]
\centering
\includegraphics[width=12cm] {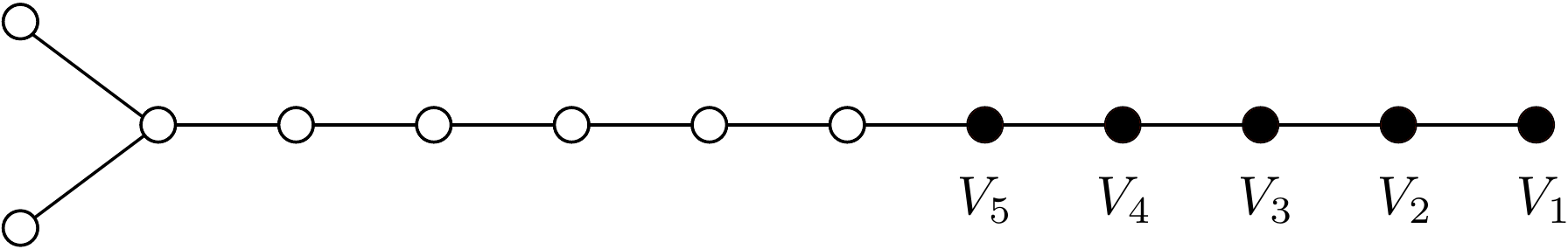}
\caption{Dynkin Diagram of the $U$-duality group $SO(5,21)$, which encodes the non-perturbative moduli space ${\cal M}_{(5,21)}$,
as per ref.~\cite{Aspinwall:1995td}.
 The black dots denote five non-compact directions in which one can take large distance limits.
 The directions $V_1$ and $V_2$ correspond to the standard limits of weak 10d dilaton and large volume, respectively.
 Note that the basis of the $V_i$ is not the same as the basis of $S_i$ which correspond to the weak coupling limits of 2-forms that we consider.
}
\label{f:SO5-21}
\end{figure}

The possible non-compact limits of the moduli space ${\cal M}_{(5,21)}$ 
in (\ref{modulispaceK31}) have been analyzed in \cite{Aspinwall:1995td} in terms of splitting
the Dynkin diagram of $SO(5,21)$ into pieces by removing the respective relevant black dots \cite{Witten:1995ex}
(see Figure \ref{f:SO5-21}). This analysis mostly focused on the resulting moduli spaces but not on 
tensionless strings that may arise in the various limits.
The point of our work is to analyze the appearance of tensionless strings from the viewpoint of dually weakly coupled tensionless Type IIB strings,
in agreement with the idea of emergence. 

%-------------------------------------------------------------
%-------------------------------------------------------------

For two out of these five possible weak coupling limits,  the appearance of tensionless strings
 is evident, because these limits are nothing but the standard limits for the six-dimensional 2-form tensor fields that descend from the various ten-dimensional
 tensor fields.  More specifically,
let us fix our conventions by starting from the ten-dimensional action in the Einstein frame, with $\ell_s \equiv 1$,
\be
S^{\rm E}_{\rm 10d} = 2 \pi \int_{\mathbb R^{1,9}} \left(\sqrt{- |g|} R  - \frac{1}{2}   e^{-\Phi}  dB_2 \wedge \ast d B_2  -     \frac{1}{2} e^{\Phi}   dC_2 \wedge \ast d C_2 - \frac{1}{4} d C_4 \wedge \ast d C_4 + \ldots \right) \,,
\ee
in terms of which the tensions of the various branes read
\bea \label{tensioncomp1}
T_{Dp} &=& 2\pi  e^   {\Phi \,  ( \frac{p+1}{4} - 1 ) } \,, \qquad \quad  \Phi: \text{10d Einstein frame dilaton} \\
T_{F1} &=&  2 \pi   e^{ \Phi/2  } \,, \qquad  \quad \quad \quad  T_{NS5} =  2\pi  e^{  - \Phi/2  } \label{tensioncomp2} \,.
\eea
Then, after compactifying on a K3 surface with volume ${\cal V}$, the relevant part of the action is
\bea \label{kinterms6dE}
S^{\rm E}_{\rm 6d} &=&\int_{\mathbb R^{1,5}} \left(\frac{1}{2} M^4_{\rm Pl} \sqrt{- |g|} R  -  \frac{2 \pi}{4}  (S_2^2 dB_2 \wedge \ast d B_2 +     S_2^{-2} d \tilde B_2 \wedge \ast d \tilde B_2) \right.   \\
 && \left. \qquad  - \frac{2 \pi}{4}  (S_1^2 dC_2 \wedge \ast d C_2 +     S_1^{-2} d \tilde C_2 \wedge \ast d \tilde C_2) + \ldots  \right)
% \frac{1}{2}   e^{-\Phi}  dB_2 \wedge \ast d B_2 +    \frac{1}{2} e^{\Phi}   dC_2 \wedge \ast d C_2 + \frac{1}{4} d C_4 \wedge \ast d C_4 + \ldots \right) \,,
\eea
with
 \be
M^4_{\rm Pl} = 4 \pi {\cal V} \,, \qquad S_2 =  e^{-\Phi/2} \sqrt{\cal V} \,, \qquad S_1 = e^{\Phi/2} \sqrt{\cal V} \,.
\ee
We have expressed the kinetic terms in a democratic fashion by employing the dual fields $\tilde B_2$ and $\tilde C_2$, which arise from the
ten-dimensional six-form fields, $B_6$ and $C_6$, and which satisfy:
\be
\ast dB_2 = S^{-2}_2  \, d \tilde B_2 \,, \qquad \quad {} \ast dC_2 = S^{-2}_1 \, d \tilde C_2  \,. 
\ee
The properly normalised tensions\footnote{Alternatively we can preform a 6d Weyl rescaling to bring the Einstein-Hilbert term into the form $\frac{1}{2} \int_{\mathbb R^{1,5}} \sqrt{-g} R$. The resulting rescaling of the string tensions corresponds to division by $M^2_{\rm Pl}$ in (\ref{F1D1tensions}). Note furthermore that $\Phi$ refers to the dilaton in the 10d Einstein frame.} of the $F1$ and $D1$ strings and of their magnetic duals,  $\tilde F1$ and $\tilde  D1$, are then
\bea \label{F1D1tensions}
\frac{T_{F1}}{M^2_{\rm Pl}} &= \sqrt{\pi} \frac{e^{\Phi/2}}{\sqrt{\cal V}} \equiv  \sqrt{\pi} \, S_2^{-1} \, \qquad &\frac{T_{\tilde F1}}{M^2_{\rm Pl}}  = \sqrt{\pi} \frac{\sqrt{\cal V}}{e^{\Phi/2}} \equiv  \sqrt{\pi} \, S_2\,, \\
\frac{T_{D1}}{M^2_{\rm Pl}} &= \sqrt{\pi} \frac{1}{ e^{\Phi/2} \sqrt{ \cal V}} \equiv  \sqrt{\pi} \, S_1^{-1}      \, \qquad &\frac{T_{\tilde D1}}{M^2_{\rm Pl}}  = \sqrt{\pi} e^{\Phi/2} {\sqrt{ \cal V} } \equiv      \sqrt{\pi} \, S_1 \,.
\eea
Thus $S_2 \to \infty$ is the weak coupling limit for the six-dimensional field $B_2$ (or strong coupling limit for its dual $\tilde B_2$), as follows from the kinetic terms in (\ref{kinterms6dE}).
Consistently, the fundamental string ${\cal S}_2 \equiv F1$ becomes asymptotically tensionless, while its ``magnetically'' dual string,  $\tilde F1$,
 becomes heavy and decouples.
On the other hand, the tension of the $D1/\tilde D1$ string system, compared to the  $F1/\tilde F1$ string system, depends on the priori independent value of 
\be \label{defV1}
e^{\Phi} \equiv \frac{S_1}{S_2} =: V_1\,.
\ee
If we keep $S_1$ fixed and finite, we have that
$e^{{\Phi}} \equiv \frac{S_1}{S_2} \ll 1$ as $S_2 \to \infty$, and so the F1 string is indeed the lightest string and can be properly viewed as the fundamental Type IIB string.

Conversely, in the weak coupling limit of the field $C_2$, where $S_1 \to \infty$, the string  ${\cal S}_1 \equiv D1$ becomes light and eventually tensionless, while its dual $\tilde D1$ decouples.  As long as we are in the geometric regime, where  the volume
\be  \label{defV2}
{\cal V} \equiv S_1 \,  S_2 =: V_2
\ee
is fixed and large, the D1-string is furthermore parametrically lighter than the string $\tilde F1$, while its dual, $F1$, decouples.

The two limits, $S_1\to \infty$ or $S_2\to \infty$, are of course S-dual to each other, in the sense of ten-dimensional $SL(2,\mathbb Z)$ symmetry:  The ten-dimensional  S-duality transformation maps $\Phi \to -\Phi$ in the 10d Einstein frame dilaton, and hence acts as interchange
\be \label{10dSduality}
\text{ 10d S-duality:} \qquad   S_1 \leftrightarrow S_2 \,.
\ee

It is important to note that in order to properly define infinite distance limits, one must precisely specify in which way variables become large, 
ie., how their ratios behave. Mathematically this is tied to the notion of a proper compactification of the moduli space, $\bar{{\cal M}}_{(5,21)}$.
In the present work, we consider limits for which the volume of K3 stays fixed in order to keep gravity dynamical, which is of importance for our investigation of
the Swampland Distance Conjectures for 2-form fields. This amounts to considering the limits $S_i\to \infty$, as opposed to products thereof such as (\ref{defV2}),
as it is the $S_i$ which canonically couple to the 2-form fields. 

As is evident from (\ref{defV1}) and (\ref{defV2}), this basis of non-compact 
moduli is different from the basis of the $V_i$ which are tied to the eponymous
nodes in the Dynkin diagram
in Figure \ref{f:SO5-21}.  The $V_i$ figure in the discussion \cite{Aspinwall:1995td} of how the moduli space ${\cal M}_{(5,21)}$ splits up upon taking 
non-compact limits by removing dots in the Dynkin diagram. 
These limits are different to the less singular limits in the $S_i$ that we take, and correspond to
a different compactification of the moduli space.

More precisely, in the infinite distance limit where $V_1 =S_1/S_2 \to 0$, we recover the weakly coupled perturbative string on K3, whose moduli space is the moduli space  of  $N=(4,4)$ supersymmetric sigma-models on K3. The latter is given by ${\cal M}_{(4,20)} =O(4,20) /  O(4) \times O(20)$ (modulo discrete identifications), which is obtained from the full moduli space (\ref{modulispaceK31}) by removing the node $V_1$ in the Dynkin diagram of Figure \ref{f:SO5-21}  \cite{Aspinwall:1995td}.
Similarly, for $V_2= S_1 S_2  \to \infty$, one obtains the 10d Type IIB string compactified on a large volume K3, whose moduli space contains ${\cal M}_{(3,19)\times(1,1)}=O(3,19) /  (O(3) \times O(19)) \times SU(1,1)/U(1)$. The first factor is the geometric moduli space of a K3 surface of fixed volume and the second is the non-perturbative moduli space of the eight-dimensional Type IIB string. Note that this large volume limit is compatible with the $SL(2,\mathbb Z)$ action (\ref{10dSduality}). 
\subsection{Geometric weak coupling limits in K\"ahler moduli space of K3}

The upshot of the previous section is that two of the five possible weak coupling limits of Type IIB string theory on K3 have a straightforward explanation 
that involves only combinations of the 10d dilaton $e^{\Phi}$ and the overall volume ${\cal V}$ of K3. 
The remaining three weak coupling limits must correspond to non-trivial degeneration limits of the K3 surface as such. 
We will analyze one of these in detail now, and make the relation between the weak coupling regime and the appearance of a solitonic tensionless string, which will turn out to be a  Type IIB string again.

The remaining weak coupling limits must involve three combinations of the  
 $22$ tensor fields $B^A$ from decomposing $C_4$ as in (\ref{C4decomp}), with kinetic terms
 \be
 S_{\rm kin} =  - \frac{2\pi}{4} \int_{\mathbb R^{1,5}} g_{AB}   dB^A \wedge \ast dB^B
  \ee
 governed by 
 \be \label{gABform}
g_{AB} =  \int_{K3}   \omega_A \wedge \ast \omega_B \,.
\ee
It is convenient to pick an arbitrary complex structure on K3 and consider the resulting Hodge decomposition of $H^2(K3,\mathbb R)$.
For any such chosen complex structure, the Hodge numbers of K3 are
\be \label{hodgenumbers}
h^{(2,0)}(K3)=1 \,, \qquad h^{(1,1)}(K3) = 20 , \qquad h^{(0,2)}(K3)=1 \,.
\ee
%where the distinction between the $(2,0)$ and $(1,1)$ forms refers to a given choice of complex structure. 
Let us specify a basis of 2-forms  by
\be
\{ \omega_A \}    = \{\Omega, \bar\Omega, \omega_\alpha \} \,,  \qquad  \omega_\alpha  \in H^{(1,1)}(K3) \,, \qquad \Omega \in H^{(2,0)}(K3) \,.
\ee
Here $\Omega$ and $\bar\Omega$ denote the unique $(2,0)$ form and its complex conjugate.

The space $H^2(K3,\mathbb R)$ can also be decomposed into spaces of self-dual and  anti-self-dual 2-forms with respect to the Hodge star operator on K3,
\bea
&&H^2(K3,\mathbb R)  =H^2_+(K3,\mathbb R)  \oplus H^2_-(K3,\mathbb R) \,,   \\
 && {\rm dim} H^2_+(K3,\mathbb R)  = 3 \,, \, \qquad {\rm dim} H^2_-(K3,\mathbb R) =19 \,.
\eea
The space of self-dual 2-forms is spanned by the real and imaginary parts of $\Omega$ and the K\"ahler form $J$. 
According to the discussion of the previous section, the tensors which become weakly coupled in the three remaining weak coupling limits 
must involve linear combinations of the self-dual tensors associated with reduction of 
$C_4$ along these three directions in $H^2_+(K3,\mathbb R)$, together with three anti-self-dual tensors form expansion of $C_4$
along three directions in $H^2_-(K3,\mathbb R)$.

In the given complex structure, one of the weak coupling limits therefore involves only a tensor field and its dual associated with the $(1,1)$ forms on K3.
It is this weak coupling limit on which we focus in the sequel.\footnote{The remaining two infinite distance limits necessarily involve complex structure deformations, which are beyond the scope of our analysis.}
On the space $H^{1,1}(K3)$, 
the matrix of kinetic terms of the 2-forms $B^\alpha$ takes the form
\be \label{kinmatK3}
g_{\alpha \beta} =  \frac{1}{\cal V} j_\alpha j_\beta - \Omega_{\alpha \beta}\,,
\ee
where $J  =  j^\alpha \omega_\alpha$ and $\Omega_{\alpha \beta} = \omega_{\alpha} \cdot \omega_\beta$.
The weak coupling limit for the system of tensor-multiplet 2-forms is therefore characterized by demanding that the kinetic matrix $g_{\alpha \beta}$ has at least one entry which tends to infinity. 

Note that the K\"ahler form implicitly depends on the choice of complex structure with respect to which  the Hodge type of the 2-forms is defined.
In this sense, the matrix $g_{\alpha \beta}$ depends on all geometric moduli of the K3 and not only on the moduli entering $J$. 
A related peculiarity of K3 surfaces is that the rank of the Picard group
\be
{\rm Pic}(K_3) = H^{1,1}(K3) \cap H_2(K3,\mathbb Z) 
\ee
depends on the choice of complex structure. 
The elements of the Picard group describe the curve classes $H_2(K3,\mathbb Z)$ which have a holomorphic representative. 

A simplification occurs when the rank of ${\rm Pic}(K_3)$ takes its maximal value: ${\rm rk}{\rm Pic}(K_3)=20$. 
This means that all curve classes have holomorphic representatives, and their volume reduces to the K\"ahler volume with respect to the K\"ahler form $J$ of the K3 surface, viewed as a 
K\"ahler manifold with vanishing anti-canonical bundle $\bar K = 0$. Such K3 surfaces are referred to as {\it singular K3s} in mathematics, and as {\it attractive K3s} in the physics literature \cite{Moore:1998pn}.
Attractive K3 surfaces lie dense in the period domain of K3 surfaces, in a similar way as $\mathbb Q^n$ lies dense in $\mathbb R^n$.
The complex structure of an attractive K3 is completely fixed. In particular, the kinetic matrix (\ref{kinmatK3}) depends solely on the K\"ahler moduli of the attractive K3.
We will for now focus on such K3 surfaces.

On an attractive K3, the weak coupling limit for (\ref{kinmatK3}) 
takes a completely analogous form as for the F-theory two-fold bases, which were discussed in section \ref{sec_emergenthet}.
Indeed in order for at least one eigenvalue of (\ref{kinmatK3}) to become large, the K\"ahler form $J$ of the compact K3 must asymptote to
\be \label{WCL-K3}
J = t J_0 + \sum_i \frac{a_i}{2 t} J_i \qquad t \to \infty \,,   \qquad \text{subject to} \, (\ref{J0J0zero}).
\ee
In the basis 
\be \label{C^Idef-K3}
\{C^I\} = \{C^0:= J_0, \quad C^i := J_i\} 
\ee
the matrix of kinetic terms takes the asymptotic form (\ref{g001}) and (\ref{g002}), and the unique 2-form potential which becomes weakly coupled is 
precisely $B^0$. It couples to the solitonic string that arises from wrapping a D3-brane on the holomorphic curve $C^0$.  The latter has the property  
\be
C^0 \cdot C^0 = 0 \,,
\ee
and moreover its volume vanishes in the weak coupling limit,
\be
{\rm vol}(C^0) = \int_{C^0} J = \frac{\cal V}{t} \qquad {\rm as} \quad t \to \infty \,.
\ee
Indeed, since the K\"ahler cone generators are dual to the Mori cone of effective curves, they are by themselves integral (1,1) classes, and on an attractive K3 every such class has a  representative as a holomorphic curve.
The novelty for this limit for K3, as compared to a K\"ahler base $B_2$ with non-trivial anti-canonical bundle, is that the genus of $C^0$ equals
\be \label{genusone}
g(C^0) = 1 + \frac{1}{2}(C^0 \cdot C^0 - C^0 \cdot \bar K) = 1 \,.
\ee
The existence of a class $C^0 \in H^{1,1}(K3) \cap H_{2}(K3,\mathbb Z)$ with vanishing self-intersection implies that the K3 surface is a genus-one fibration, with fiber $C^0$, i.e. there exists a projection
\bea \label{K3ellfibered}
\pi :\quad C^0 \ \rightarrow & \  \ K3 \cr 
& \ \ \downarrow \cr 
& \ \  C_b \,.
\eea
See e.g. \cite{Huybrechts} for a proof of this theorem, which is the analogue of Kollar's conjecture for K3 surfaces.\footnote{This can be seen as a physical proof for the geometrical statement that attractive K3 surfaces are genus-one fibered, under the assumption that weak coupling limits exist in the K\"ahler moduli space. Indeed, it is an established fact in mathematics that a K3 surface with Picard number $\rho$ is genus-one fibered if $\rho\geq 5$ and is elliptic with a section if $\rho\geq 13$; see e.g.~\cite{SS}.}

We can summarize these findings by stating that, in the notation of Section \ref{sec_weakcouplSUGRA}, the tensor field 
\be
{\rm T}_3 := B^0
\ee
 is singled out as the tensor which is  weakly coupled in the geometric limit $t \to \infty$.
Its kinetic terms can asymptotically be written as 
\be  \label{SkinT3}
S_{\rm kin} = -  \frac{2\pi}{4} \int_{\mathbb R^{1,5}}(S^2_3 d B^0 \wedge \ast d B^0 + S^{-2}_3 d \tilde B^0 \wedge \ast d \tilde B^0)
\ee
with
\be
S_3 = \frac{t}{\sqrt{\cal V}} \,.
\ee
If we also introduce the notation
\be
{\cal S}_3 := {\rm String \, \,  from \, \, a\, \,  D3-brane \, \, on} \,  C^0 \,,
\ee
then the normalised tension of ${\cal S}_3$ scales as
\be \label{S3tension}
T_{{\cal S}_3}/M^2_{\rm Pl} = \sqrt{\pi} \,  \frac{\sqrt{\cal V}}{t} =: \sqrt{\pi} \, S^{-1}_3 \,.
\ee

In (\ref{SkinT3}) we have exhibited also the magnetically dual field $\tilde B^0\equiv \tilde{\rm T}_3$.
Correspondingly the solitonic string ${\cal S}_3$ has a magnetic dual $\tilde {\cal S}_3$ arising from a D3-brane wrapping a dual cycle on K3. This cycle can be obtained in principle by carefully going through the procedure outlined at the end of section \ref{weakcouplinghet}: We need to bring the duality matrix associated with the kinetic matrix (\ref{g00gijasmp}) obtained in the limit into the form (\ref{B-basis}), which conforms with (\ref{SkinT3}). Let us denote the cycle associated with $\tilde B^0$ as $C''^1$, as in section  \ref{weakcouplinghet}.
Even without determining its cycle class explicitly we anticipate from the expected tension of the dual string that its volume must scale as
\be
{\rm vol}_{\tilde C''^1} \sim t \,.
\ee
Clearly such cycles are present in the Picard lattice: In the limit (\ref{WCL-K3}) this is the behaviour of a cycle with 
\be
J_0 \cdot C''^1 \neq 0 \,.
\ee
This  string becomes heavy and strongly coupled in the limit $t \to \infty$ for fixed ${\cal V}$, and effectively decouples.

%%%%%%%%%%%%%%%%%%%%%%%%%%%%%%%%%%%%%%%%%%%%%%%%%%%
\subsection{The solitonic string as a  Type IIB fundamental string}
%%%%%%%%%%%%%%%%%%%%%%%%%%%%%%%%%%%%%%%%%%%%%%%%%%%

% We now consider a D3-brane that wraps such a genus-one curve $C^0$. In the limit  $\to \infty$ it gives rise to a weakly coupled string
%  with asymptotically vanishing tension
% \be \label{TC0IIB}
% T_{C^0} = 2 \pi \, {\rm vol}(C^0) = \frac{\cal V}{t} \qquad {\rm as} \quad t \to \infty \,.
% \ee
The tensionless string ${\cal S}_3$, in fact, describes a critical, fundamental Type IIB string propagating on $\mathbb R^{1,5} \times K3$.
This follows from (\ref{genusone}) in view of the findings of \cite{Katz:1999xq}, which we now briefly review from the perspective of our purposes.

The worldsheet theory of ${\cal S}_3$ is derived via dimensional reduction of the worldvolume theory on a single D3-brane wrapped on the genus-one curve $C^0$. The latter is, of course, given by 4d $N=4$ Super-Yang-Mills theory with gauge group $U(1)$. 
Compactification on $C^0 \subset K3$ breaks one half of the 16 supercharges on the D3-brane and results in a string worldsheet theory with $N=(4,4)$ supersymmetry.

This can be made manifest by performing a standard topological twist along $C^0$: The $SU(4)_R$ symmetry of the D3-brane theory, which is identified with the $SO(6)_T$ rotation group acting on the dimensions transverse to the D3-brane, decomposes as
\be
SO(6)_T \rightarrow SO(4)_T \times U(1)_R \,, \qquad      SO(4)_T  =  SU(2)_L \times SU(2)_R \,,
\ee
where $SO(4)_T$ acts on the four extended directions $\mathbb R^4_T$ normal to the string. The Lorentz group parallel to the D3-brane decomposes into
\be
SO(1,3) \rightarrow SO(1,1) \times U(1)_C \,,
\ee
where the two factors act on the extended and internal dimensions of the string, respectively.
The topological twist is defined with respect to the combination
\be
U(1)_{\rm tw} = \frac{1}{2} (U(1)_C + U(1)_R) \,.
\ee
Only the supercharges which are singlets under this $U(1)_{\rm tw}$ are preserved.
Twisted dimensional reduction of the $N=4$ vector multiplet leads to one $N=(4,4)$ hypermultiplet along with one $N=(4,4)$ vector multiplet, as displayed in Table \ref{table_(4,4)}.\footnote{More details can be found e.g. in section 3.1.1 of \cite{Lawrie:2016axq}. This reference performs, in addition to the standard topological twist, a duality twist, which must be omitted for K3 with trivial canonical bundle.}

\begin{table}
  \centering
  \begin{tabular}{|c|c|c|c|c|c|}
    \hline
      \multicolumn{2}{|c|}{Fermions} &
    \multicolumn{2}{c|}{Bosons}     & $N=(4,4)$ &  Multiplicity \\ \hline\hline
     ${\bf (2,1)}_1 + c.c.$  &$\psi_{+} + c.c.$& ${\bf (1,1)}_0$, ${\bf (1,1)}_0$
    & $ \bar{a}, \bar{\sigma}$ &\multirow{2}{*}{Hyper} & \multirow{2}{*}{ $h^{0,1}(C^0) = g=1$ }      \\
 ${\bf (1,2)}_{-1} + c.c.$ &$\tilde{\rho}_{-} + c.c. $   & ${\bf (1,1)}_0$, ${\bf
    (1,1)}_0$ & $a,\,\sigma$ & &   \\ \hline \hline
    %  ${\bf (1,2)}_{-1} + c.c.$ &$\tilde{\rho}_{-} + c.c. $& & &  &   \\
    % ${\bf (1,2)}_{-1}$ & ${\rho}_{-}$& &&&   \\ \hline\hline 
      ${\bf (1,2)}_1 + c.c.$ & $\mu_+ + c.c. $& \multirow{1}{*}{${\bf (2,2)}_{0}$} & \multirow{1}{*}{$\varphi$} &  \multirow{2}{*}{ Vector  } & \multirow{2}{*}{$h^{0,0}(C^0) = 1$} \\
      ${\bf (2,1)}_{-1} + c.c.$ & $\lambda_{-} + c.c.$&     ${\bf (1,1)}_{\pm 2}$    &  $v_\pm$   & &     \\ \hline\hline

    %  ${\bf (2,1)}_{-1}$ & $\lambda_-$& ${\bf (1,1)}_{2}$& $v_+$ &\multirow{2}{*}{Vector} &\multirow{2}{*}{$h^1(C,K_C \otimes {\cal L}_D) = 0$}\cr
    %  ${\bf (2,1)}_{-1}$ & $\tilde{\lambda}_- $&${\bf (1,1)}_{-2}$ & $v_-$ &&\cr\hline
  \end{tabular}
  \caption{Spectrum of massless 2d $N=(4,4)$ multiplets on the worldsheet of a string obtained by wrapping a D3-brane on the genus-one curve $C^0 \subset K3$ with $C^0 \cdot C^0=0$.
  The first column indicates the representations with respect to $SU(2)_L \times SU(2)_R \times SO(1,1)$.  We denote by $+/-$ the 2d left/right-moving chiralities. \label{table_(4,4)}}
  \end{table}

Following the arguments of \cite{Bershadsky:1995vm}, in the limit of vanishing curve volume for $C^0$ the gauge kinetic term for the $U(1)$ potential $v_\pm$ along the string
decouples and the worldsheet theory reduces to a non-linear sigma model. 
If we ignore the non-dynamical vector field $v_\pm$, the field content agrees precisely with that of a critical, fundamental Type II string propagating on $\mathbb R^{1,5} \times K3$: The four real scalars $\varphi$ transform as a vector under $SO(4)_T$
and describe the fluctuations in the transverse directions $\mathbb R^4_T \subset \mathbb R^{1,5}$. Together with their superpartners they are associated with a free sector of the worldsheet theory. 
The remaining four real scalars  $a$, $\bar a$, $\sigma$, $\bar \sigma$ and their superpartners are the fundamental fields of an interacting non-linear sigma model. Two of these, $a$ and $\bar a$, are associated with the string motion along $C^0$, while $\sigma$ and $\bar \sigma$ describe the fluctuations in the internal directions normal to $C^0$. 

The target space of the of non-linear sigma model is identified with the moduli space of the string  \cite{Bershadsky:1995vm}.
Apart from the external directions transverse to the string, this moduli space coincides precisely the original manifold K3 \cite{Katz:1999xq}.
To see this, one makes use of the fact that the wrapped curve $C^0$ is the elliptic fiber of K3, according to (\ref{K3ellfibered}). 
The moduli space of $C^0$ as a holomorphic curve is identical to the base $C_b$ over which it is fibered. In addition, the sigma-model moduli space includes the moduli space of flat 
gauge backgrounds on a D3-brane on $C^0$. The latter are described by the Jacobian of the curve, which gives back the curve itself (or rather its dual). As a result, one recovers the full fibration of $C^0$ over $C_b$ as the internal part of the moduli space of the string on $C^0$ \cite{Katz:1999xq}. Together with the external part this yields a non-linear sigma-model on $\mathbb R^{4}_T \times K3$.

%%%%%%%%%%%%%%%%%%%%%%%%%%%%%%%%%%%%%%%%%%%%%%%%%%%%
\subsection{Emergent strings and duality in the geometric weak coupling limit} \label{subsec_emergent}
%%%%%%%%%%%%%%%%%%%%%%%%%%%%%%%%%%%%%%%%%%%%%%%%%%%%

The previous considerations make it evident that, as we take the large distance, weak coupling limit (\ref{WCL-K3}), we can 
switch to a new duality frame, denoted by hatted quantities in the sequel. The solitonic string 
${\cal S}_3$ (D3 brane wrapped on $C_0$) in the original frame turns into the weakly coupled, {\it fundamental Type IIB string} $\widehat F1 \equiv \widehat {\cal S}_2$ propagating on $K3$.
The weakly coupled 2-form field ${\rm T}_3 := B^0$ with coupling $S_3$ given in (\ref{S3tension}) in the old frame takes the role of the fundamental field $\widehat {\rm T}_2 \equiv \hat B_2$ in the new frame, with corresponding 'dilaton' $\widehat S_2 = S_3$:
%\be
%\widehat F1  = \text{D3 wrapped on} \,   C_0 \,.
%\ee
\begin{equation}
\boxed{
\label{hatgs}
\text{Type IIB on K3}    \quad  \stackrel{\text{limit} \, (\ref{WCL-K3})}{\Longrightarrow}    \quad   \text{Type IIB   on K3 with}    \,     \widehat S^2_2 := {\cal V} \, e^{-\widehat \Phi} = \frac{t^2}{\cal V} }
% \end{center}
\end{equation}
% \ee
As in the analogous F-theory/heterotic correspondence we are identifying the volume of the original theory with the volume of the new theory to the extent that we insist on keeping the Planck scale fixed.

In the duality frame defined by the string ${\cal S}_3 = \widehat {\cal S}_2$, there must also arise a heavy S-dual string  $\widehat D1 \equiv \widehat {\cal S}_1$.
Its associated coupling $\widehat S_1$ follows from the coupling $\widehat S_2 = S_3$ by demanding that 
\be
\widehat S_1   \widehat S_2 = \widehat{\cal V} \equiv {\cal V} 
\ee
in analogy to the standard relation (\ref{defV2}) for the S-dual tensions in the weakly coupled fundamental frame.
Together with (\ref{hatgs}) this gives
\be
\widehat S_1^{-1} = \frac{t}{{\cal V}^{3/2}} \,.
\ee 
Such a string can arise from a D3-brane wrapping a curve $\Sigma$ on K3 with
\be \label{volSigma}
{\rm vol}(\Sigma) = \frac{t}{\cal V} \,.
\ee
The magnetic dual of this string, $\widehat{\tilde D}_1 \equiv \widehat{\tilde {\cal S}}_1$ has normalised tension
\be
\frac {T_{\widehat{\tilde{\cal S}}_1}}{M^2_{\rm Pl}}  = \sqrt{\pi} \widehat S_1  =  \sqrt{\pi}  \, \frac{{\cal V}^{3/2}}{t} = {\cal V} \,   \frac {T_{\widehat{{\cal S}}_2}}{M^2_{\rm Pl}}      \, ,
\ee
which vanishes in the limit $t \to \infty$, even though it is enhanced compared to the tension of the new fundamental string ${\cal S}_3\equiv \widehat{\cal S}_2$ by a factor of ${\cal V}$.

It is interesting to wonder which curve $\Sigma$ can give rise to the behaviour (\ref{volSigma}).
All curves in the Picard lattice obtain their volume exclusively from the intersection with the K\"ahler form $J$.
As pointed out already, the volumes of such curves which become large in the limit $t \to \infty$ are the ones with non-zero overlap with the K\"ahler cone generator $J_0$, but their volumes scale as $t$ as opposed to $t/{\cal V}$. This argument seems to suggest that the curve $\Sigma$ obtains its volume rather from the overlap with the $(2,0)$ form $\Omega$, in the sense of a symplectic integral.
A natural conjecture is therefore that the S-dual curve lies in the overlap of $(H^{2,0}(K3) \oplus H^{0,2}(K3)) \cap H_2(K3,\mathbb Z)$, or at least receives contributions from this space. 
Note that for attractive K3s, this space is 2-dimensional \cite{Aspinwall:2005ad}.

An important point to keep in mind is that the moduli space of six-dimensional theories with $N=(2,0)$ supersymmetry is not corrected by worldsheet and D1-instanton effects.
Hence even though Euclidean F1 and D1 strings (of the original duality frame) wrapped on the vanishing cycle(s) on K3 might look like giving rise to a non-suppressed instanton effect, they will not affect the classical couplings in the limit we are considering. The magnetic duals of such instantons in six dimensions are (3+1)-dimensional objects in $\mathbb R^{1,5}$ and correspond to the original D5 or NS5-brane wrapping curves on K3. For the vanishing curve $C^0$, the tension of these objects scales as $ T_{3+1} \sim {\rm vol}(C^0) = {\cal V}/t$ 
and hence vanishes at the same time as the tension of the new fundamental string $\widehat{\cal S}_2$ goes to zero.
However, the associated mass scales compare as
\be
M_{3+1} = T_{3+1}^{1/4}   \sim \left(\frac{\cal V}{t}\right)^{1/4} \gg \left(\frac{\cal V}{t}\right)^{1/2} \sim    T_{\widehat {\cal S}_2}^{1/2} =  M_{\widehat {\cal S}_2}    \Longrightarrow \frac{M_{3+1}}{M_{\widehat {\cal S}_2}} \sim \left(\frac{\cal V}{t}\right)^{-1/4} \to \infty \,.
\ee
Hence these objects decouple in the duality frame defined by the new fundamental string $\widehat {\cal S}_2$.

Even though the string ${\cal S}_3$ takes the role of the new fundamental string $\widehat S_2$ on K3, the limit $t \to \infty$ at fixed ${\cal V}$, i.e. $S_3 \to \infty$ at fixed ${\cal V}$, is different from
the weak coupling limit $S_2 \to \infty$ at fixed ${\cal V}$ studied in section \ref{sec_weakcouplSUGRA}. 
The reason is that in addition to the tower of excitations from ${\cal S}_3$ which becomes massless, there is a tower of asymptotically light Kaluza-Klein (KK) states which arises because we are considering Type IIB on a K3 in a very special deformation limit. 
% The only other source of potentially
% light particle states is the tower of Kaluza-Klein excitations,
%  which inevitably arise whenever some directions on the compactification space $K3$ become very large.
The origin of the KK tower are the 
cycles which become large in the limit (\ref{WCL-K3}), i.e. all curves with $C_{\rm large} \cdot J_0 \neq 0$. Since their volume scales as  ${\rm vol}(C_{\rm large}) \sim t$,
the mass scale of the Kaluza-Klein excitations associated with these curves 
is 
% \footnote{The same conclusion formally holds for the scale of the winding strings around the 1-cycles of the shrinking curve $C_0$,
% \be
% M_{\rm winding} \sim {\rm vol}(C_0)^{1/2} = \frac{1}{t^{1/2}} = \widehat g_s^{1/4} \sim   M_{\widehat F1} \,.
% \ee
% However, since these 1-cycles do of course not survive as 1-cycles on K3, the winding strings can slip off, and no such winding objects arise. }
\be
M^2_{\rm KK} \sim {\rm vol}(C_{\rm large})^{-1} = \frac{1}{t}  \sim   \frac{T_{{\cal S}_3}}{{\cal V}} \equiv  \frac{T_{\widehat{\cal S}_2}}{{\cal V}} \,.
\ee
 The states with mass scale $M_{\rm KK}$ are the Kaluza-Klein excitations of the new fundamental string $\widehat{\cal S}_2$ in the new Type IIB frame.
This is consistent in that $\widehat{\cal S}_2$ arises from a D3-brane wrapped on the elliptic fiber $C^0$ of K3, and it therefore makes sense to consider the Kalzua-Klein spectrum of its excitations, at each excitation level, along the remaining directions on K3. 

Let us contrast the limits $S_3 \to \infty$ versus $S_2 \to \infty$ as follows:

\bea
&&{\rm Limit} \, \, S_2 \to \infty, \,  {\cal V} \gg 1  \, \,{\rm fixed}: \qquad e^{\Phi} \to 0  \label{limitS2a}  \\
&& (M^2_{{\cal S}_2}  = e^{\Phi/2}) \quad   <  \quad   (M^2_{\tilde{\cal S}_1}  = e^{\Phi/2} {\cal V})  \quad  < \quad (M^2_{KK} = \frac{1}{\sqrt{\cal V}}) \quad    <  \quad (M^2_{\rm Pl} = \sqrt{\cal V})   \nonumber \\
&&{\rm Limit} \, \, S_3 \to \infty, \,  {\cal V} \gg 1 \, \,{\rm fixed}: \qquad \frac{1}{t} \to 0   \label{limitS3a}\\
&& (M^2_{\rm KK}  = \frac{1}{t}) \quad   <  \quad   (M^2_{\widehat{\cal S}_2}  = \frac{\cal V}{t})  \quad  < \quad (M^2_{\widehat{\tilde S}_1} = \frac{{\cal V}^2}{t}) \quad    <  \quad (M^2_{\rm Pl} = \sqrt{\cal V})   \nonumber
\eea

The interesting feature of the geometric weak coupling limit (\ref{limitS3a}) is that for fixed ${\cal V}$, the KK excitations sit at the same parametric
scale as the excitations of $\widehat {\cal S}_2$ itself, as far as the scaling with the weak coupling parameter $t$ is concerned.
In particular, as the new string tension $T_{\widehat {\cal S}_2} = \frac{2\pi {\cal V}}{t}$ asymptotes to zero, so does the KK scale, albeit relatively suppressed by a factor of ${\cal V}$.
Hence the limit $t\to \infty$ governs both the asymptotically massless tower of oscillator excitations of the new fundamental weakly coupled string, $\widehat {\cal S}_2$, as well as simultaneously its Kaluza-Klein tower of states. In both situations (\ref{limitS2a}) and (\ref{limitS3a}) there arises in addition the tower of the magnetic version of the S-dual string, which is enhanced by another factor of ${\cal V}$. 
Despite the appearance of the KK tower in the limit (\ref{limitS3a}) we would not call this a decompactification limit in the usual sense: Such a limit is detected by a encountering {\it only} a tower of light KK states, without an accompanying tower of extra string modes.

The observation that the tower of massless states here includes the tower of a fundamental string sheds some light on the proposal of emergence  put forward in \cite{Harlow:2015lma,Heidenreich:2017sim,Heidenreich:2018kpg,Grimm:2018ohb,Palti:2019pca}:
According to the general lore, in the weak coupling limit at infinite field distance, a tower of states becomes  massless exponentially fast \cite{Ooguri:2006in}; integrating out this tower of  states leads to
the running of the coupling constants of the theory and reproduces the polynomial singularity at infinite distance in moduli space, as encountered in the effective supergravity description.
 Evidence for this proposal has been provided in \cite{Harlow:2015lma,Heidenreich:2017sim,Heidenreich:2018kpg,Grimm:2018ohb,Lee:2018urn,Corvilain:2018lgw} by estimating, at a {\it qualitative} level, the contribution of the integrated tower of particle states to the renormalization of the couplings.
On the other hand, an {\it exact} computation reproducing the observed supergravity coupling constants near infinite distance has not been obtained in the literature, as it is a priori much more difficult to quantify the spectrum of the asymptotically massless states. 
 
 What we are observing here is that the massless tower of states has a very clear interpretation as exactly the excitations of the critical fundamental string $\widehat {\cal S}_2$ in the new duality frame plus its 
 Kaluza-Klein excitations. 
 Importantly, the new fundamental string is again the Type IIB string probing the same K3 (in the specific geometric limit).
  Integrating out the entire tower of string excitations together with the KK modes reproduces, {\it by construction}, the coupling dependence of the original supergravity theory in the weak coupling limit.
  This is true not only parametrically, but at an exact level, by the very definition of the low-energy effective supergravity as the consequence of integrating out the heavy string states and the Kaluza-Klein/winding modes on the internal space.
 In this sense, even without explicitly evaluating this integration, it is manifest that the polynomial divergence in the coupling constants at infinite distance is exactly reproduced and, in fact, caused by the tower of states which become asymptotically massless.

 %%%%%%%%%%%%%%%%%%%%%%%%%%%%%%%%%%%%%%%%%%%%%%%%%
\section{Emergent F-theory from M-theory on K3 in Weak Coupling Limit} \label{sec_emergentF}
 %%%%%%%%%%%%%%%%%%%%%%%%%%%%%%%%%%%%%%%%%%%%%%%%%

As we have shown in the previous sections, following refs.~\cite{Lee:2018urn,Lee:2018spm}, 
weak coupling limits in the K\"ahler moduli space of F-theory and Type IIB compactifications lead to emergent critical, nearly tensionless strings along with their towers of oscillator and Kaluza-Klein excitations. As we will show in this section, the same geometrical limits in K\"ahler moduli that we have analyzed above
 have a strikingly different effect as probed by M-theory or Type IIA string theory. A detailed analysis of infinite distance limits in K\"ahler moduli space from the perspective of Type IIA/M-theory has been provided before in \cite{Corvilain:2018lgw}, and we will comment on the relation of this work to our findings below.

Consider thus M-theory compactified on a K3 surface to d=7 dimensions, which leaves $16$ supercharges unbroken. 
The crucial difference to the situation in F-theory or Type IIB theory is that  this theory contains
 $h^2(K3) = 22$ $1$-form gauge fields, as opposed to 2-form fields, 3 of which are part of the gravity multiplet. %and we can envisage taking a weak coupling limit for one or several of the remaining $19$ vector fields, $A^\alpha$. 
These $1$-form fields are obtained by the reduction of the M-theory 3-form $C_3$:
\be
C_3 = A^\alpha \wedge \omega_\alpha \, \qquad \omega_\alpha \in H^{1,1}(K3) \,.
\ee
The same coupling matrix of kinetic terms as in (\ref{kinmatK3}) now applies
to the 1-form gauge fields $A^\alpha$, as opposed to the 2-forms $B^\alpha$ in Type IIB/F-theory. 

For simplicity, let us restrict ourselves to attractive K3 surfaces with a Picard group of maximal rank, as in the previous section. 
Then, despite the different interpretation, the analysis of the weak coupling limit at fixed 7d Planck scale proceeds in an entirely identical fashion:
It corresponds to taking the limit 
\be \label{WCL-K3-b}
J = t J_0 + \sum_i \frac{a_i}{2 t} J_i \,, \qquad t \to \infty  \,, \qquad \text{subject to} \, (\ref{J0J0zero}).
\ee
In the basis of (1,1)-forms $\{\omega_{\alpha}\} = \{ J_0, J_i \}$, the gauge field $A^0$ is the only linear combination of 1-form potentials which becomes asymptotically weakly coupled as $t \to \infty$. The pertinent gauge kinetic matrix can be found in (\ref{g00gijasmp}).
 
Thus, a unique genus one curve $C^0:= J_0$ shrinks at the rate ${\rm vol}(C^0) = \frac{\cV}{t}$.
What becomes massless in the weak coupling limit is now a tower of particles (as opposed to a string and its excitations), which arise
from M2-branes wrapping the shrinking curve $C^0$.
Importantly, the Gopakumar-Vafa invariants for an $n$-fold wrapped genus-one curve on K3 are \cite{Katz:1999xq}
\be
N^{\rm GV}_{n \cdot C^0} = 24 \, \qquad \forall n \geq 1 \,.
\ee
This fact guarantees the existence of a tower of asymptotically massless states in the effective theory.

As we have discussed around eq.~(\ref{K3ellfibered}),
in order for the K3 to admit the weak coupling limit (\ref{WCL-K3-b}), it must be fibered with a genus one curve $C^0$,
\bea \label{K3ellfibered-b}
\pi :\quad C^0 \ \rightarrow & \  \ K3 \cr 
& \ \ \downarrow \cr 
& \ \  C_b \,.
\eea
The weak coupling limit hence amounts to shrinking the genus one fiber of the K3 while keeping its total volume fixed.
This limit coincides exactly with the F-theory limit associated with M-theory on K3:
\begin{center}
\begin{minipage}{5cm}
\text{F-theory on $C_b \times S^1$ }   \vspace{1mm}\\
 $S^1$-radius $R$
\end{minipage}
\begin{minipage}{2cm}
 $\Longleftrightarrow$
\end{minipage}
\begin{minipage}{5cm}
M-theory on K3 as in (\ref{K3ellfibered-b}) \vspace{1mm}  \\
${\rm vol}(C^0) = \frac{1}{R}$
\end{minipage}
\\
\vspace{1mm}
\end{center}
Recall that F-theory on a genus-one fibered K3-surface over the base $C_b$ gives rise to an eight-dimensional 
 theory with 16 supercharges. Upon compactification on a circle $S^1$ with radius $R$, we obtain a 7d theory which is identified with M-theory on the same K3, in the limit where the fiber volume scales as $\frac{1}{R}$.
Each 8d field maps to a full Kaluza-Klein tower of excitations in the 7d M-theory, which are in turn interpreted as M2-branes wrapping the fiber $C^0$ $n$-times.
These M2-branes are charged under the Kaluza-Klein gauge potential $A_{\rm KK}$.

We therefore see that the weak coupling limit for an  - a priori arbitrary  - 1-form gauge field in M-theory on K3,  inevitably enforces the F-theory limit:
The asymptotically weakly coupled gauge field, $A^0$ is identified with the Kaluza-Klein $U(1)_{\rm KK}$ gauge group in the F-theory/M-theory correspondence, and the tower of asymptotically massless particles from M2-branes along the curve $n \cdot C^0$ represent the Kaluza-Klein tower associated with the 8d supergravity modes of F-theory. 

 Hence, we encounter an emergent extra dimension in the weak coupling limit, in which the duality frame of the theory switches from that of 7d M-theory to the duality frame of 8d F-theory:
\begin{equation}
\boxed{
\text{7d M-theory on K3} \qquad \stackrel{ \text{limit} (\ref{WCL-K3-b}) }{\Longrightarrow} \qquad \text{8d F-theory on} \, C_b }
\end{equation}
\vspace{1mm}

%The remaining 19 gauge fields uplift to 18 gauge fields in F-theory due to the presence of 7-branes, plus a 2-form gauge potential arising from the reduction of the 4-form $C_4$ on $C_b$. 
\noindent Importantly, the gauge fields in the F-theory are not in a weak coupling regime. Rather the weakly coupled M-theory 1-form has become part of the 8d metric, since it refers to the $U(1)$ gauge field associated with the Kaluza-Klein reduction.

Note that infinite distance limits in K\"ahler moduli space and their interpretation in Type IIA and M-theory have been analyzed in detail in \cite{Corvilain:2018lgw}.
This reference includes a study of infinite distance limits on elliptically fibered Calabi-Yau 3-folds, and, as a special case of such limits, the F-theory limit of vanishing fiber volume.
The key point of our discussion is that whenever one considers a weak coupling limit in the K\"ahler moduli space, on an attractive K3, the shrinking curve is automatically a genus one fiber of the K3, and the weak coupling limit reduces to the F-theory limit, even without making any assumption about the fibration structure of the K3 surface.
In this sense the emergence of the extra dimension in F-theory is a consequence of the weak coupling limit.

%%%%%%%%%%%%%%%%%%%%%%%%%%%%%%%%%%%%%%%%%%%%%%%%%%%
\section{Conclusions, and Prospects for Four-dimensional Strings}  \label{sec_concl}
%%%%%%%%%%%%%%%%%%%%%%%%%%%%%%%%%%%%%%%%%%%%%%%%%%%

In this work we have explored the behavior of effective theories near infinite distance points in moduli space where $2$-form gauge fields become weakly coupled, in the presence of gravity.  Specifically we have systematically analyzed such weak coupling limits  for six-dimensional compactifications of F-theory and Type IIB theory, with $N=(1,0)$ and $N=(2,0)$ supersymmetry, respectively. Two universal phenomena have emerged in the limit: 
\begin{itemize}
\item [1)] A critical string becomes tensionless, be it the fundamental string in the original duality frame or, generically, a solitonic string which takes the role of the fundamental string in a dual frame.
\item [2)] An infinite tower of asymptotically massless particles arise as the excitations of the emergent critical string,  in agreement with the Swampland Distance Conjecture. \end{itemize}
% which verify the $2$-form version of the Weak Gravity Conjecture and the Swampland Distance Conjecture, respectively. 

For F-theory compactifications, we have obtained the general form of the weak coupling limits of $2$-forms.
The limit turns out to be identical to the weak coupling limits of $1$-form gauge fields studied in~\cite{Lee:2018urn}, in the K\"ahler moduli space of the internal complex surface. 
By the same arguments as in ~\cite{Lee:2018urn} a solitonic string emerges as a critical, weakly coupled heterotic string.

% The same argument as in this latter work has thus lead to the identification of an internal curve class $C^0$ with its volume asymptoting to zero. It has turned out that this curve class is uniquely determined (modulo scaling) amongst all the classes with a non-negative self-intersection. Furthermore, it has to have a trivial normal bundle and hence corresponds to a rational curve. Therefore, $C^0$, when wrapped by a D3 brane, leads to an emergent tensionless string, which we have identified as a weakly coupled heterotic string in the dual heterotic frame. As a consequence, this critical heterotic string upon its quantization gives rise to an infinite tower of particles that become masselss in the limit. 

On the other hand, for Type IIB compactifications, the moduli space has five different types of non-compact directions.
 Two obvious directions correspond to the limits where the ten-dimensional Kalb-Ramond and Ramond-Ramond $2$-forms are weakly coupled, and hence where the fundamental string or the D1 string becomes tensionless, respectively. Amongst the remaining three `geometric' weak coupling limits, we have systematically analyzed the one reachable within the K\"ahler moduli space of the internal K3 surfaces, upon restricting to geometries with maximal Picard number $\rho=20$. We have singled out a unique elliptic curve class  with non-negative normal bundle
whose volume vanishes in the weak coupling limit.
  % It has turned out, as in the F-theory case, that the class $C^0$ has a trivial normal bundle. Upon its embedding into a K3 surface, 
A D3 brane wrapping this curve leads to an emergent tensionless string, which we have identified as a weakly coupled Type IIB string in yet another Type IIB duality frame. Its quantum excitations give rise to an infinite tower of massless particles in the limit. The remaining two non-compact weak coupling limits lie in the complex structure moduli space and we leave their further investigation to future work. 

An interesting aspect of weak coupling limits in the K\"ahler moduli space, for both F-theory and Type IIB compactifications, is that another set of particles become asymptotically massless, in addition to the tower arising from the quantization of the relevant, nearly tensionless string. These are the Kaluza-Klein modes that become light in the geometric limits we consider.
%where there must exist an expansing curve as well as the shrinking one $C^0$. This is an inevitable consequence of fixing the volume of the internal space, which is required to keep the gravity dynamical. 
The masses of these KK modes exhibit the same suppression factor by the large K\"ahler parameter as the string modes, although an additional suppression by
the volume, $\cV$, of the K3 is inevitable. 
Since the KK tower becomes massless at the same rate as the string tower,
the limit should not be interpreted as the unfolding of a new dimension. We have contrasted this situation with the weak coupling limits of $1$-form potentials in M-theory compactifications on K3 surfaces, once again restricting the discussion to  geometries with maximal Picard number. In such limits, the fiber class of an elliptic fibration  shrinks, uplifting the seven-dimensional M-theory compactification to  F-theory in eight dimensions. 

While this work has focused on effective theories in $d=6$ dimensions,
many of our conclusions pertaining the nature of tensionless strings carry over to four dimensions, as we now briefly discuss:
For the four-dimensional compactifications of both F-theory and Type IIB theory, the $2$-form fields $B^I$ arise from expanding $C_4$ over a basis $C_I$ of $H^{1,1}(B_3)$ and have the kinetic coupling matrix
\be
g_{IJ} = - \frac{1}{4 \cV} C_I \cdot C_J \cdot J + (\frac{1}{4\cV})^2 (C_I \cdot J \cdot J)   (C_J \cdot J \cdot J) \,, 
\ee 
where $\cV=\frac{1}{6}\int_{B_3} J^3$ is the volume of the internal three-fold $B_3$. This $B_3$ is the base of an elliptic four-fold in F-theory,  while on the other hand it coincides
with the internal Calabi-Yau three-fold when we talk about a Type IIB string compactification. In the weak coupling limit, at least one entry of $g_{IJ}$ diverges (and hence at least one of the K\"ahler parameters goes to infinity) while $\cV$ is kept fixed. Geometric limits of this kind have been considered in Section 6.1 of~\cite{Lee:2019tst} and classified as limits of Class A and Class B, respectively. 

In the weak coupling limits of Class A, the K\"ahler form takes the following non-negative expansion in terms of the K\"ahler cone generators,
\be \label{Alimit1}
J = t J_0 + \sum_\nu \frac{a_\nu}{t^2} J_\nu + \sum_r c_r J_r \,, \quad\quad\text{with~~} J_0 \cdot J_0 \neq 0\,, \quad t \to \infty\,,
\ee
where $J_0 \cdot J_0\cdot J_0 = 0$ and $J_0 \cdot J_0 \cdot J_\nu >0$ for $\nu$ in a certain index set, while $J_0 \cdot J_0 \cdot J_r =0$ and $J_0 \cdot J_r \cdot J_s=0$ for $r$ and $s$ in another index set. It was then proven that there exists a curve $C^0:=J_0 \cdot J_0$ with a trivial normal bundle which shrinks in such a geometric limit. 
For F-theory compactifications, as argued in~\cite{Lee:2019tst}, such a curve has to be a rational curve and hence a D3 brane wrapping $C^0$ plays the role of the tensionless critical heterotic string in the heterotic duality frame.

 If $B_3$ is itself  a Calabi-Yau three-fold, on the other hand,  the existence of the K\"ahler cone generator $J_0$ implies \cite{Kollar, Oguiso, Wilson}  that 
$B_3$ is necessarily genus-one fibered;  the shrinking curve $C^0$ is the fiber of this fibration. 
 The world-sheet theory associated with a D3-brane wrapped on $C^0$ can be determined by methods similar to those spelled out in \cite{Lawrie:2016axq}: The result is a world-sheet theory with $N=(2,2)$ supersymmetry.  Its bosonic excitations include two real scalars parametrizing the motion of the string in the two extended directions transverse to the string, two complex scalars transforming as sections of the normal bundle $N_{C^0/B_3} = {\cal O} \oplus {\cal O}$ to the wrapped curve and one complex scalar transforming as a section of ${\cal O}_{C^0}$, corresponding to the Wilson line degrees of the freedom. The moduli space of the D3-brane along $C^0$ is the  fibration of $C^0$ over the base of the fibration, and hence coincides with $B_3$. In the limit of shrinking curve volume we therefore manifestly arrive at a sigma-model with target space $B_3$, which we interpret as the critical Type IIB string propagating on $B_3$.  By the same logic as in Section~\ref{sec_emergentIIB} of the present paper, 
 we conclude that the emerging tensionless string is the Type IIB string in another Type IIB duality frame. 
The theory, again, reproduces itself in the dual weak coupling limit at infinite distance.

It is intriguing to see that  the Calabi-Yau three-fold is necessarily elliptic in order to admit a weak coupling limit of the form (\ref{Alimit1}).
For five-dimensional M-theory compactifications,
the existence of such a geometric limit corresponds to a weak coupling limit for $1$-form potentials.
As for M-theory on K3 analysed in this paper,
we can immediately conclude  that a weak coupling limit of Class A for 5-dimensional M-theory is a decompactification limit which coincides with 
 the F-theory uplift to six dimensions. This is a consequence of the properties of the limit (\ref{Alimit1}) and holds  without making any a priori assumptions about the existence of an elliptic fibration.
 
 The situation is more complicated for the weak coupling limits of Class B \cite{Lee:2019tst}, however, which we leave to future work. 

In view of these results in the context of F-theory and Type IIB theory, it is natural to wonder  about the 
nature of the tensionless strings that emerge in weak coupling limits for $2$-forms in Type IIA compactifications. Six-dimensional Type IIA compactifications on K3 surfaces do not have $2$-forms which can become  weakly coupled in the geometric moduli space. On the other hand, in four-dimensional Type IIA compactifications on Calabi-Yau three-folds $2$-forms arise from expanding either the Ramond-Ramond $5$-forms over harmonic $3$-forms or the dual $6$-form of the Kalb-Ramond $2$-form over harmonic $4$-forms. These respectively admit weak coupling limits in the complex structure and K\"ahler moduli space \cite{Grimm:2018ohb,Grimm:2018cpv,Corvilain:2018lgw,Font:2019cxq} of the Calabi-Yau three-folds. 
Mirror symmetry suggests that analogous phenomena as in the Type IIB case should occur. 
The pressing question in this context is whether the resulting nearly tensionless strings are dual to some weakly coupled strings, similarly to what
we have found here for six-dimensional compactifications of F-theory and Type IIB theory.

\subsection*{Acknowledgements}

We thank Yang-Hui He, Fernando Marchesano, Eran Palti, Christian Reichelt, Fabian R\"uhle, Max Wiesner and Fengjun Xu for helpful discussions. The work of SJL is supported by the Korean Research Foundation (KRF) through the CERN-Korea Fellowship program.

% \newpage
\bibliography{papers}
\bibliographystyle{JHEP}

\end{document}